\documentclass[aps,prl,twocolumn,10pt]{revtex4-2}
\usepackage{amssymb}
\usepackage{graphicx}
\usepackage{amsmath}
\usepackage{hyperref}
\usepackage{subfigure}
\usepackage{multirow}
\usepackage{setspace}
\usepackage{verbatim}
\usepackage{float}
\usepackage{color}
\usepackage{ulem}
\usepackage[utf8]{inputenc}

\begin{document}

\title{First detection of the Hubble variation correlation and its scale dependence}

\author{Wang-Wei Yu$^{1,2}$}
\author{Li Li$^{1,2,3,4}$}
\author{Shao-Jiang Wang$^{1}$}
\email{Corresponding author: schwang@itp.ac.cn}

\affiliation{$^1$CAS Key Laboratory of Theoretical Physics, Institute of Theoretical Physics, Chinese Academy of Sciences, Beijing 100190, China}
\affiliation{$^2$School of Physical Sciences, University of Chinese Academy of Sciences (UCAS), Beijing 100049, China}
\affiliation{$^3$School of Fundamental Physics and Mathematical Sciences, Hangzhou Institute for Advanced Study (HIAS), University of Chinese Academy of Sciences (UCAS), Hangzhou 310024, China}
\affiliation{$^4$Peng Huanwu Collaborative Center for Research and Education, Beihang University, Beijing 100191, China.}
%\date{\today}

\begin{abstract}
The sample variance due to our local density fluctuations in measuring our local Hubble-constant ($H_0$) can be reduced to the percentage level by choosing the Hubble-flow type Ia supernovae (SNe Ia) outside of the homogeneity scale. In this Letter, we have revealed a hidden trend in this one-percent $H_0$ variation both theoretically and observationally. We have derived for the first time our $H_0$ variation measured from any discrete sample of distant SNe Ia. We have also identified a residual linear correlation between our local $H_0$ fitted from different groups of SNe Ia and their ambient density contrasts of SN-host galaxies evaluated at a given scale. We have further traced the scale dependence of this residual linear trend, which becomes more and more positively correlated with the ambient density contrasts of SN-host galaxies estimated at larger and larger scales, on the contrary to but still marginally consistent with the theoretical expectation from the $\Lambda$-cold-dark-matter model. This might indicate some unknown corrections to the peculiar velocity of the SN-host galaxy from the density contrasts at larger scales or the smoking gun for the new physics.
\end{abstract}
\maketitle

\textbf{Introduction.}---
The Hubble constant $H_0$ measures the current background expansion rate of our observable Universe. Although $H_0$ is not one of six base parameters of the $\Lambda$-cold-dark-matter ($\Lambda$CDM) model, its derived value is crucial in establishing a concordant cosmology among different observations~\cite{Moresco:2022phi}. The most stringent value $H_0=67.27\pm0.60$ km/s/Mpc comes from globally fitting the $\Lambda$CDM model to the cosmic microwave background (CMB) data of Planck 2018 results~\cite{Planck:2018vyg}. However, local measurements from type Ia supernovae (SNe Ia) calibrated by Cepheids ~\cite{Riess:2016jrr,Riess:2018byc,Riess:2018uxu,Riess:2019cxk,Riess:2020fzl} favor significantly higher values in tension with CMB constraints. This Hubble tension~\cite{Bernal:2016gxb,Verde:2019ivm,Knox:2019rjx,Riess:2020sih,DiValentino:2020zio,DiValentino:2021izs,Perivolaropoulos:2021jda,Abdalla:2022yfr} seems to be a crisis~\cite{Schoneberg:2021qvd,Jedamzik:2020zmd,Cai:2021weh,Cai:2022dkh} since the most recent measurement  $H_0=73.04\pm1.04$ km/s/Mpc~\cite{Riess:2021jrx} with an unprecedented~$\sim5\sigma$ discrepancy. A comprehensive list of analysis variations considered to date have been thoroughly investigated to contribute insignificantly to the total error budget, offering no particularly promising solution to the Hubble tension.

As the systematic errors from the external photometric calibration have been persistently reduced over the years, recent renewed focus~\cite{DES:2022tgg,Wojtak:2022bct,Rose:2022zmu,Dixon:2022ryo,DES:2022qsy,DES:2022zpw,Jones:2022tsf} has been shifted to the physical origin of the intrinsic scatter of standardized SN Ia brightnesses. The most pronounced variation in the standardized  SN Ia brightness comes from an ad-hoc step-like correction as a function of the host-galaxy stellar mass, which, as a global property of the SN-host galaxy, is hardly directly related to the SN itself, yet strongly correlated to the distance-modulus residuals. This host-galaxy stellar mass correlation~\cite{Kelly:2009iy,SNLS:2010kps,SDSS:2010swx,Gupta:2011pa,Johansson:2012si,Childress:2013xna} to the Hubble residual remains elusive as a long-standing puzzle over the past decade, during which considerable efforts~\cite{NearbySupernovafactory:2013qtg,Rigault:2014kaa,Jones:2015uaa,Uddin:2017rmc,Roman2018,Jones:2018vbn,Rose:2019ncv,Brout:2020msh,Popovic:2021cwq} have been made towards possible interpretations as a result of Hubble residual correlations to other global and local properties of SN-host galaxy. 

%In particular, a recently suggested dust-based model~\cite{Brout:2020msh} has attributed the mass step to the differences in dust properties for SNe with different colors, though it is still inconclusive whether the recovered dust properties are intrinsic or extrinsic to the SN.

Perhaps the most global property of SN-host galaxy is its local  density contrast. It is well-known that~\cite{1992AJ....103.1427T,Shi:1995nq,Shi:1997aa,Wang:1997tp} our local $H_0$  measurement can be deviated from the global value due to our own local density contrast with its standard deviation decreasing with an enlarging sample volume. Therefore, to reduce this sample variance to the percentage level, a redshift range $0.023<z<0.15$~\cite{Sinclair:2010sb,Marra:2013rba,Ben-Dayan:2014swa,Camarena:2018nbr} is usually adopted for the sample selection to obviate the effects from a large local density contrast around us and the dark energy at higher redshift. Indeed, our local density contrast has been checked to be incapable of accounting for the Hubble tension~\cite{Wojtak:2013gda,Odderskov:2014hqa,Wu:2017fpr,Kenworthy:2019qwq,Lukovic:2019ryg,Cai:2020tpy,Castello:2021uad,Camarena:2022iae}. However, the effect on our local $H_0$ measurement from the local density contrasts of the SN-host galaxies has never been explored before. This can be motivated from the aforementioned Hubble residual correlation to the stellar masses of SN-host galaxies that usually populate in the denser regions for more massive halos than less massive halos~\cite{Sheth:2000ii}.

In this Letter, we have theoretically derived for the first time our local $H_0$ variation from arbitrary discrete sample of distant SNe Ia. A hidden trend is then revealed by fitting our local $H_0$ from different groups of SNe Ia selected in such a way that they share the same value for their ambient density contrasts estimated at a given scale. Increasing this scale leads to a more and more positive correlation between the local $H_0$ values fitted from different groups of SNe Ia and their corresponding ambient density contrasts of SN-host galaxies, which is in direct contrast to but still marginally consistent with our theoretical estimation from the $\Lambda$CDM model. This might be caused by unknown systematics or new physics. We stress that the current study is not aimed at solving the Hubble tension, or explaining the mass step correction, or testing for a large local void, but revealing a new Hubble-constant variation correlation in SN data.

\begin{figure*}[htbp]
\includegraphics[width=0.45\textwidth]{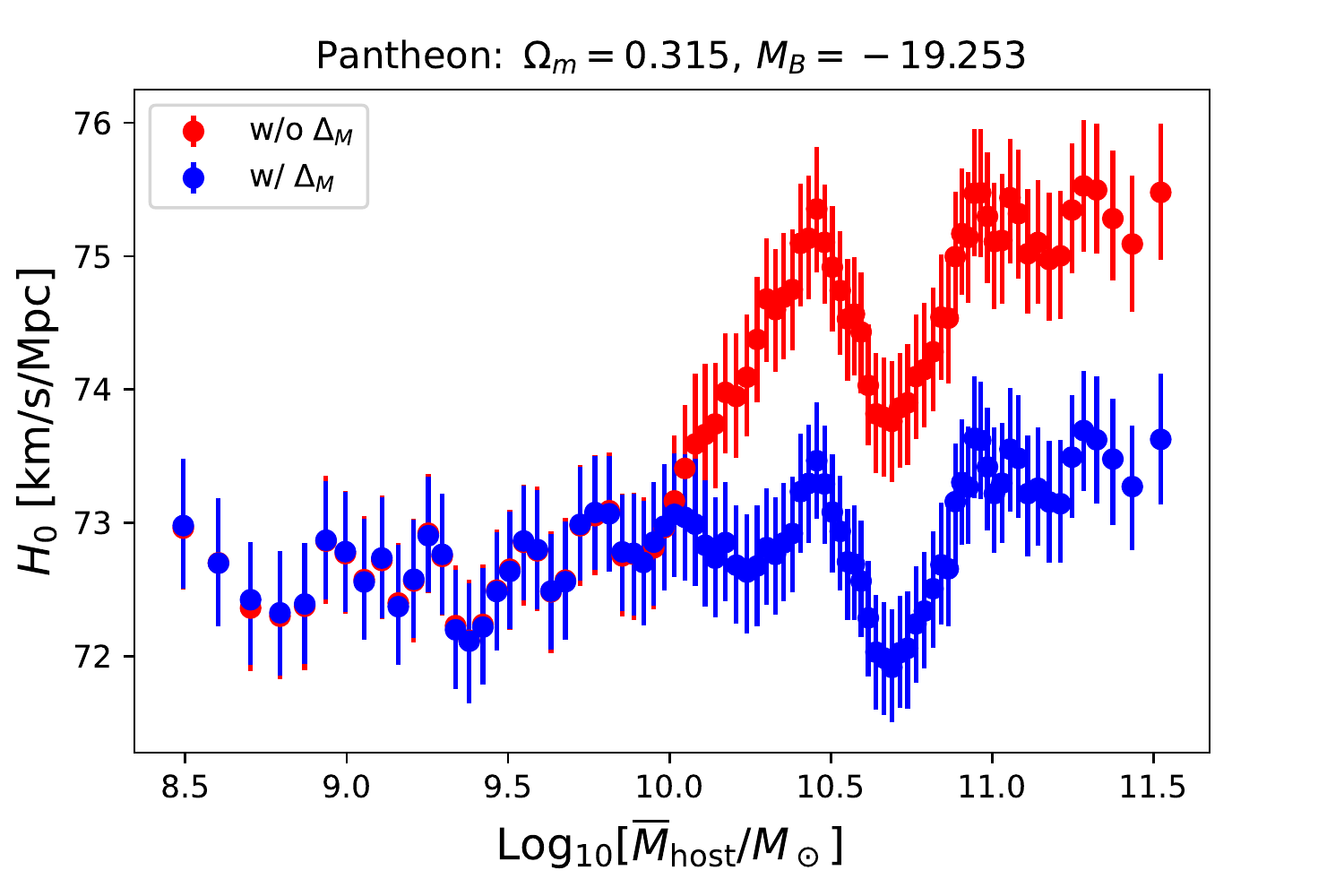}
\includegraphics[width=0.45\textwidth]{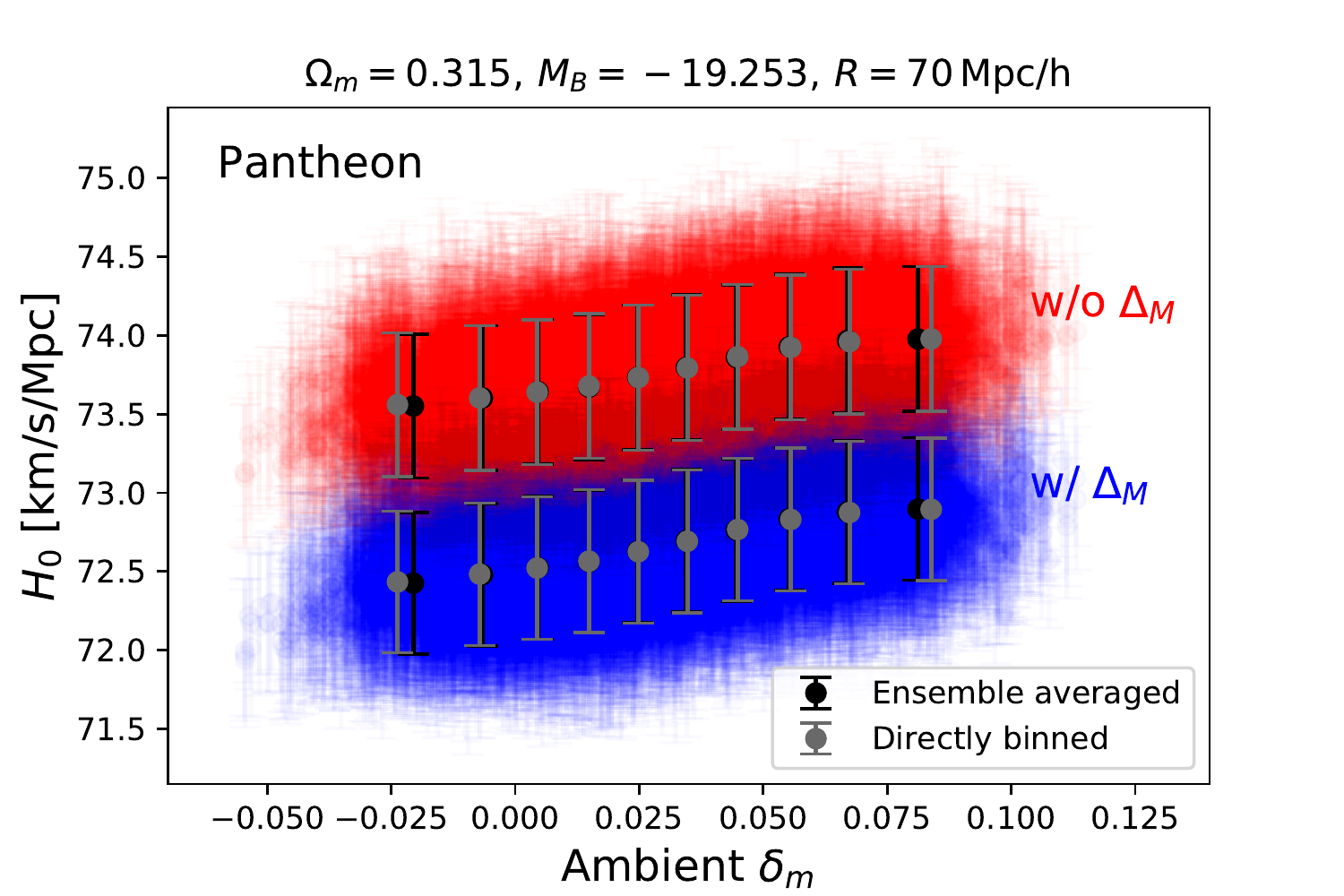}\\
\caption{The host-galaxy stellar-mass (left) and ambient-density (right) correlations to our local $H_0$ values (with $1\sigma$ error bars) fitted from different groups of Pantheon SNe with respect to the averaged host-galaxy stellar-mass (left) and ambient density (right) of each group. In the right panel, the extra black/gray points with error bars are the field-averaged/direct-binned $H_0$ values since the ambient densities of each SN-host galaxy are estimated from 2000 reconstructed density fields at an illustrative smoothing scale $R=70\,\mathrm{Mpc}/h$. In both panels, the inclusion and exclusion of the mass step correction are indicated in blue and red, respectively, and $\Omega_\mathrm{m}=0.315$ and $M_B=-19.253$ are fixed for illustration.}\label{fig:HostCorrelation}
\end{figure*}

\textbf{Host-galaxy stellar mass correlation.}---
It has long been known~\cite{Kelly:2009iy,SNLS:2010kps,SDSS:2010swx,Gupta:2011pa,Johansson:2012si,Childress:2013xna} that the SNe Ia appear to be intrinsically fainter in the host galaxies with higher stellar masses than those with lower stellar masses. Therefore, SNe Ia in high mass galaxies would be more luminous after the standardization corrections than those in low mass galaxies. To account for this host mass correlation in the observed distance modulus,
\begin{align}
\mu_\mathrm{obs}=m_B-M_B+\alpha x_1-\beta c+\Delta_M+\Delta_\mathrm{bias},
\end{align}
a step-like correction $\Delta_M$ is usually added by hand in addition to corrections due to the stretch $x_1$ and color $c$ as well as predicted biases $\Delta_\mathrm{bias}$ from simulations. Here $m_B$ is the observed  B-band peak magnitude, and $M_B$ is the absolute B-band magnitude of a fiducial SN Ia with $x_1=0$ and $c=0$ obtained externally from local distance ladder calibrations. On the other hand, the distance modulus can also be modeled theoretically as
\begin{align}
\mu_\mathrm{mod}=5\lg\frac{D_L(z)}{10\,\mathrm{pc}}=5\lg d_L(z)+5\lg\frac{\langle c\rangle}{\langle H_0\rangle}+25,
\end{align}
where $\langle c\rangle$ is the value of the speed of light $c$ in the unit of km/s and $\langle H_0\rangle\equiv 100h$ is the value of $H_0$ in the unit of km/s/Mpc.  Here $d_L(z)=D_L(z)/(c/H_0)\equiv(1+z)\int_0^z\mathrm{d}z'/E(z')$ is the dimensionless luminosity distance evaluated at, for example, the $\Lambda$CDM model with $E(z)\equiv H(z)/H_0=[\Omega_\mathrm{m}(1+z)^3+1-\Omega_\mathrm{m}]^{1/2}$ approximated at the late time from the current matter fraction $\Omega_\mathrm{m}$. Then, the Hubble residual is defined as 
\begin{align}
\Delta\mu\equiv\mu_\mathrm{obs}-\mu_\mathrm{mod}\equiv m_B^\mathrm{cor}-m_B^\mathrm{mod},
\end{align}
where $m_B^\mathrm{cor}\equiv m_B+\alpha x_1-\beta c+\Delta_M+\Delta_\mathrm{bias}$ and $m_B^\mathrm{mod}\equiv5\lg d_L(z)+\mathcal{M}_B$ with $\mathcal{M}_B\equiv M_B+5\lg(\langle c\rangle/\langle H_0\rangle)+25$.

If the mass step correction is not included, then the binned Hubble residual would admit a decreasing trend with respect to the host mass, which could be more visible by directly looking at the $H_0$ values fitted from different groups of SNe Ia by their host masses. For the Pantheon sample~\cite{Scolnic:2017caz,Jones:2017udy}, the mass step correction reads
\begin{align}
\Delta_M=\gamma\times\left[1+\exp\left(-\frac{m_\mathrm{host}-m_\mathrm{step}}{\tau}\right)\right]^{-1}.
\end{align}
Here $\gamma=0.054\pm0.009$ is a relative offset in luminosity, $m_\mathrm{host}\equiv\lg M_\mathrm{host}/M_\odot$ and $m_\mathrm{step}\equiv\lg M_\mathrm{step}/M_\odot$ are in logarithmic scales, and $m_\mathrm{step}=10.13\pm0.02$ is a mass step for the split. The exponential transition term measured by $\tau=0.001\pm0.071$ describes the relative probability of masses being on one side or the other of the split to allow for uncertainties in the mass step and host masses. 
For the Pantheon+ sample~\cite{Scolnic:2021amr,Brout:2022vxf},  the mass step correction takes the form $\Delta_M\to\Delta_M-\gamma/2$ with $\gamma=0.06$, $m_\mathrm{step}=10$, and $\tau=0.001$~\cite{Peterson:2021hel}.
We adopt $\chi^2$-test by estimating
\begin{align}
\chi^2=\mathbf{\Delta\mu}^T\cdot\mathbf{C}^{-1}\cdot\mathbf{\Delta\mu}
\end{align}
with the Markov chain Monte Carlo (MCMC) code \texttt{EMCEE}~\cite{Foreman-Mackey:2012any} when fitting to  $H_0$ with a flat prior for given $M_B=-19.253$~\cite{Riess:2021jrx} and $\Omega_\mathrm{m}=0.315$~\cite{Planck:2018vyg}. Here the total covariance matrix $\mathbf{C}=\mathbf{C}_\mathrm{stat}+\mathbf{C}_\mathrm{sys}$ contains both statistical and systematic contributions~\cite{Scolnic:2017caz,Jones:2017udy}. 

The fitted $H_0$ values are shown in the left panel of Fig.~\ref{fig:HostCorrelation} for the Pantheon sample with $N=1002$ SNe Ia \footnote{we have omitted 46 SNe Ia with their hosts too faint for survey depth so that their host masses are simply assigned in the lowest mass bin} first presorted by their host masses as $\{M_n\}_{n=1}^N$ and then fitted by taking every 100 SNe out as a group each time for the MCMC analysis. The grouping strategy $N=100+(k-1)s$ is $1002=100+82\times11$ so that the $i$-th $H_0^{(i)}$ value can be fitted from the $i$-th group $\{M_{1+(i-1)s}, \cdots, M_{100+(i-1)s}\}$ with respect to its average host mass $\overline{M}_\mathrm{host}^{(i)}=(M_{1+(i-1)s}+\cdots+M_{100+(i-1)s})/100$ for $i=1,\cdots k$ with a shift $s$ between each neighbouring groups. As can be seen from Fig.~\ref{fig:HostCorrelation}, without the mass step correction, there is a overall step-like shape in the local $H_0$ values with respect to the averaged host masses.

\begin{figure*}[htbp]
\centering
\includegraphics[width=0.52\textwidth]{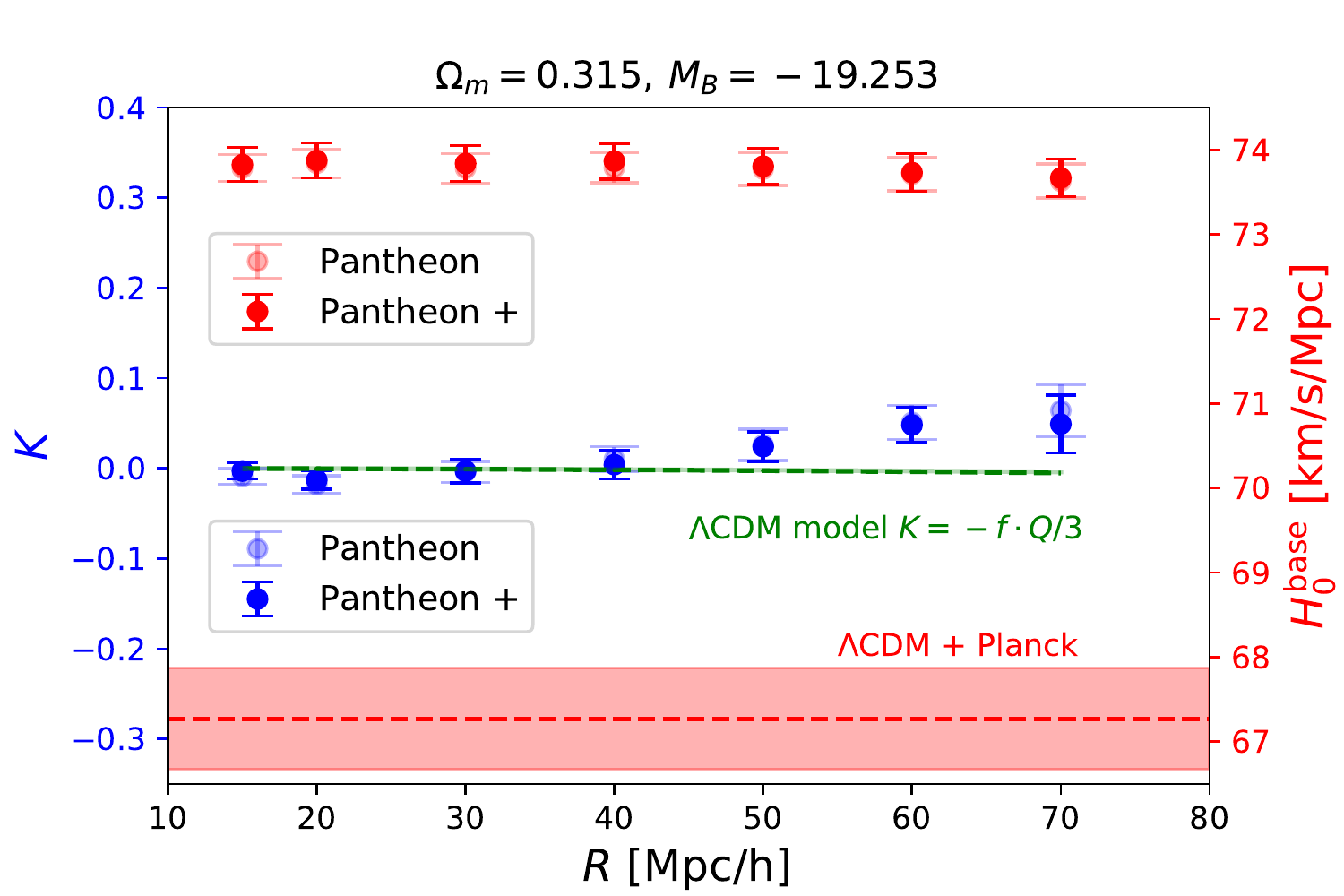}
\includegraphics[width=0.47\textwidth]{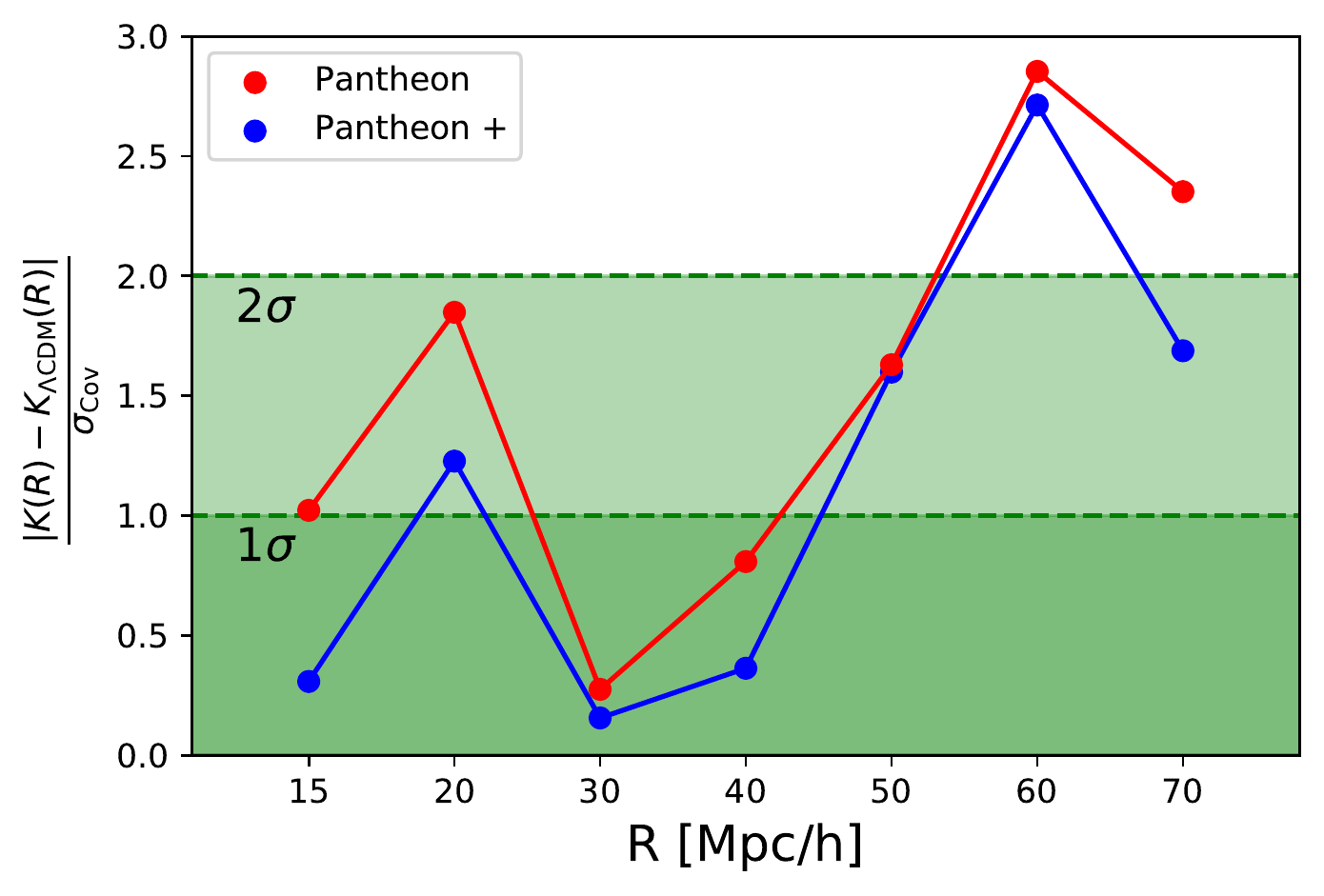}\\
\caption{\textit{Left}: The scale-dependence of the slope (blue points) and intercept (red points) fitted from a linear correlation model~\eqref{eq:ObservationalFit} to the field-averaged $H_0$ values with respect to the ambient density contrasts of SN-host galaxies at each smoothing scale. The corresponding $\Lambda$CDM prediction for the slope $K(R)$ is shown with green dashed line and the Planck constraint on $H_0$ is shown with the red band for comparison with the intercepts. \textit{Right}: The standard deviation of observational $K(R)$ with respect to the $\Lambda$CDM prediction with $1\sigma$ uncertainties estimated from diagonal elements of the inverse covariance matrix of slope correlations between different smoothing scales.}
\label{fig:ScaleDependence}
\end{figure*}

\textbf{Host-galaxy ambient density correlation.}---
The aforementioned host mass correlation motivates us to consider whether there is also an inherited correlation to the local matter density contrast of SN-host galaxy since massive halos containing SN-host galaxies with higher stellar masses prefer to populate in the denser regions than less massive halos~\cite{Sheth:2000ii}. 
This is reminiscent of the well-known sample variance~\cite{1992AJ....103.1427T,Shi:1995nq,Shi:1997aa,Wang:1997tp} for our local $H_0$ measurements affected by our local density contrast. Measuring a local Hubble expansion rate $H_0^\mathrm{loc}(\mathbf{r}_0)$ at $\mathbf{r}_0$ from a group of SNe Ia within a 3-ball $B_R^3(\mathbf{r}_0)$ of radius $R$ centered at $\mathbf{r}_0$ would result in a local variation $\delta_H(\mathbf{r}_0;B_R^3(\mathbf{r}_0))\equiv(H_0^\mathrm{loc}(\mathbf{r}_0)-H_0^\mathrm{bac})/H_0^\mathrm{bac}$ as
\begin{align}
\delta_H(\mathbf{r}_0;B_R^3(\mathbf{r}_0))=f(\Omega_\mathrm{m})\int\frac{\mathrm{d}^3\mathbf{k}}{(2\pi)^3}\widetilde{\delta}_\mathrm{m}(\mathbf{k})\widetilde{\mathcal{L}}(kR)e^{i\mathbf{k}\cdot\mathbf{r}_0}
\end{align}
on top of the background Hubble expansion rate $H_0^\mathrm{bac}$, where $f(\Omega_m)\approx\Omega_\mathrm{m}^{-0.55}$ is the linear growth rate  from the $\Lambda$CDM model, and $\widetilde{\delta}_\mathrm{m}(\mathbf{k})$ is the Fourier mode of the density contrast $\delta_\mathrm{m}(\mathbf{r})\equiv(\rho_\mathrm{m}(\mathbf{r})-\bar{\rho}_\mathrm{m})/\bar{\rho}_\mathrm{m}$ on top of the mean matter density $\bar{\rho}_\mathrm{m}$. All small-scale modes with $kR\gg1$ are integrated out by the window function $\widetilde{\mathcal{L}}(kR)\equiv[3/(kR)^3][\sin(kR)-\mathrm{Si}(kR)]$ with the sine integral $\mathrm{Si} x\equiv\int_0^x\mathrm{d}y\sin y/y$. 
The sample variance
\begin{align}
\langle\delta_H^2(\mathbf{r}_0;B_R^3(\mathbf{r}_0))\rangle=\frac{f(\Omega_\mathrm{m})}{2\pi^2R^2}\int_0^\infty\mathrm{d}k\,P_\mathrm{m}(k)[kR\widetilde{\mathcal{L}}(kR)]^2\nonumber
\end{align}
monotonically decreases with $R$ given the linear-order matter power spectrum $P_\mathrm{m}(k)\equiv\langle\widetilde{\delta}_\mathrm{m}(\mathbf{k})\widetilde{\delta}^*_\mathrm{m}(\mathbf{k})\rangle$. This sample variance can be used to select the Hubble-flow SNe outside of the homogeneity scale~\cite{Scrimgeour:2012wt} $R>R_\mathrm{homo}\approx70\,\mathrm{Mpc}/h$ (corresponding to $z>0.023$ in the $\Lambda$CDM model with $\Omega_\mathrm{m}\approx0.3$) so that the corresponding sample variance can be reduced to the percentage level~\cite{Sinclair:2010sb,Marra:2013rba,Ben-Dayan:2014swa,Camarena:2018nbr}.

On the other hand, for a sufficiently local sample  with $R\to0$, the limit $\widetilde{\mathcal{L}}(kR)\to-1/3$ would give rise to the well-known Turner-Cen-Ostriker (TCO) relation~\cite{1992AJ....103.1427T}
\begin{align}\label{eq:TCORelation}
\delta_H(\mathbf{r}_0;B_{R\to0}^3(\mathbf{r}_0))=-\frac{f(\Omega_\mathrm{m})}{3}\delta_\mathrm{m}(\mathbf{r}_0),
\end{align}
hence, an observer in a local under-dense region would always overestimate its local Hubble expansion rate. Unfortunately, such a large local void sufficiently deep to resolve the Hubble tension has been ruled out by the current observations~\cite{Wojtak:2013gda,Odderskov:2014hqa,Wu:2017fpr,Kenworthy:2019qwq,Lukovic:2019ryg,Cai:2020tpy,Castello:2021uad,Camarena:2022iae}. In particular, the TCO relation~\eqref{eq:TCORelation} can be used to define a local slope $K_\mathrm{local}\equiv\delta_H(\mathbf{r}_0; B_{R\to0}^3(\mathbf{r}_0))/\delta_\mathrm{m}(\mathbf{r}_0)=-f(\Omega_\mathrm{m})/3$ as the ratio of the Hubble-constant variation with respect to the density contrast at the same local point $\mathbf{r}_0$. This local slope is hard to be tested with the real data since it is not only subjected to the large cosmic variance from the particular choice of a local position $\mathbf{r}_0$,  but also limited to the large sample variance from a small sample volume required to be sufficiently local at $\mathbf{r}_0$.  This is why Ref.~\cite{Wu:2017fpr} can only estimate $\delta_H(\mathbf{r}_0;B_{R}^3(\mathbf{r}_0))$ from averaging $\delta_H(\mathbf{r}_i;B_{R}^3(\mathbf{r}_i))$ by positioning the observers at SN-host galaxies $\mathbf{r}_i$ in the simulation data for a non-vanishing $R\lesssim 120\,\mathrm{Mpc}/h$.

However, as we will see shortly below, there is also a hidden trend within the aforementioned percentage-level Hubble-constant variation $\delta_H(\mathbf{r}_0;B_{R>R_\mathrm{homo}}^3(\mathbf{r}_0))$ from the usual SN sample outside of the homogeneity scale as long as our local Hubble constants are fitted from different groups of SNe Ia preselected by their ambient density contrasts estimated at a given scale. In the supplemental material~\cite{footnote}, we have derived for the first time from the $\Lambda$CDM model a theoretical estimation,
\begin{align}\label{eq:HubbleVariation}
\bar{\delta}_H(\mathbf{0};\{\mathbf{r}_i|\bar{\delta}_\mathrm{m}^R(\mathbf{r}_i)=\delta_\mathrm{m}^R\})\approx-\frac{f(\Omega_\mathrm{m})}{3}\left\langle\frac{R^2}{r_i^2}\right\rangle_i\delta_\mathrm{m}^R,
\end{align}
on the variation in the measured local Hubble constant at $\mathbf{r}_0\equiv\mathbf{0}$ from an arbitrary discrete sample of distant SNe Ia at $\mathbf{r}_i$ preselected with the same averaged density contrast $\bar{\delta}_\mathrm{m}^R(\mathbf{r}_i)=\delta_\mathrm{m}^R$ over a local 3-ball centered at $\mathbf{r}_i$ of radius $R\ll R_\mathrm{homo}\ll r_i$. Different from the TCO relation~\eqref{eq:TCORelation}, this new Hubble-constant variation relation~\eqref{eq:HubbleVariation} is not only detached to the specific size and shape of the sample volume but also free from the large cosmic and sample variances. The large comic variance is evaded by pre-selecting different groups of SNe Ia with different ambient densities $\bar{\delta}_\mathrm{m}^R(\mathbf{r}_i)=\delta_\mathrm{m}^R$ at different scales. The large sample variance is absent for sufficiently distant SNe Ia at $r_i\gg R_\mathrm{homo}$. Our Hubble-constant variation correlation~\eqref{eq:HubbleVariation} also defines a non-local slope,
\begin{align}\label{eq:TheoreticalSlope}
K\equiv\frac{\bar{\delta}_H(\mathbf{r}_0\equiv\mathbf{0})}{\bar{\delta}_\mathrm{m}^R(\mathbf{r}_i)}=-\frac{f(\Omega_\mathrm{m})}{3}\left\langle\frac{R^2}{r_i^2}\right\rangle_i\equiv-\frac{f(\Omega_\mathrm{m})}{3}Q,
\end{align}
as the ratio of the Hubble-constant variation with respect to the density contrast at different points $\mathbf{r}_0\equiv\mathbf{0}$ and $\mathbf{r}_i$. We then turn to search for this host-galaxy ambient density correlation in the observational data.

\textbf{Observational search and analysis.}---
The data we adopt for estimating the ambient density contrasts of the SN-host galaxies comes from the cosmic matter density field reconstruction~\cite{Lavaux:2019fjr} from the final data release (DR12)~\cite{SDSS-III:2015hof,BOSS:2016wmc} of the Baryon Oscillation Spectroscopic Survey (BOSS). Since the density field is reconstructed from the velocity field of the galaxy tracers located at the local peaks of the underlying density field, the density reconstruction process from a galaxy survey would return back $N$ different reconstructions of the density contrast fields $\delta_\mathrm{m}^I(\mathbf{r}_i)\equiv(\rho_\mathrm{m}^I(\mathbf{r}_i)-\bar{\rho}_\mathrm{m})/\bar{\rho}_\mathrm{m}, I=1,\cdots,N$ at the reconstruction cell $\mathbf{r}_i$. We have identified $M=163$ ($M=202$) SNe Ia in the Pantheon(+) samples at positions $\mathbf{d}_i$ within the  BOSS survey volume,  which will be used for fitting the local Hubble constant with respect to different groups of SNe Ia according to their ambient density contrasts. To estimate the ambient density contrast for each selected SN at $\mathbf{d}_i$, we can average a total number $n_i$ of density field points $\mathbf{r}_j$ over a sphere of radius $R$ centered at $\mathbf{d}_i$, that is $|\mathbf{r}_j(\mathbf{d}_i)-\mathbf{d}_i|^2<R^2, j=1,2,\cdots, n_i$. Therefore, the ambient density contrast of that SN from the $I$-th ensemble can be estimated as
\begin{align}
\bar{\delta}_\mathrm{m}^I(\mathbf{d}_i)=\frac{1}{n_i}\sum_{j=1}^{n_i}\delta_\mathrm{m}^I(\mathbf{r}_j(\mathbf{d}_i)).
\end{align}

The grouping strategy is similar to that of left panel of Fig.~\ref{fig:HostCorrelation}. After assigning the ambient density contrast to each selected SN from different reconstructions of BOSS density fields, we first put all the selected SNe Ia in an ascending order $P^{I}$ as
$\bar{\delta}_\mathrm{m}^I(\mathbf{d}_{P^I_1})\leq\bar{\delta}_\mathrm{m}^I(\mathbf{d}_{P^I_2})\leq\cdots\leq\bar{\delta}_\mathrm{m}^I(\mathbf{d}_{P^I_M})$ according to their ambient density contrasts $\bar{\delta}_\mathrm{m}^I(\mathbf{d}_i)$ in the $I$-th density field. Then, we can take every 100 SNe Ia out as a group each time for fitting the $H_0$ value  $H_0(\langle\bar{\delta}_\mathrm{m}^I\rangle_k)$ with $1\sigma$ uncertainty $\sigma_{H_0}(\langle\bar{\delta}_\mathrm{m}^I\rangle_k)$ with respect to the group-averaged ambient density contrast
\begin{align}
\langle\bar{\delta}_\mathrm{m}^I\rangle_k&\equiv\frac{1}{100}\sum_{j=1}^{100}\bar{\delta}_\mathrm{m}^I(\mathbf{d}_{P^I_{j+(k-1)s}})
\end{align}
of the $k$-th group shifted by $s$ each time between two neighboring groups. Therefore, the grouping strategy $n=100+(m-1)s$ is $163=100+9\times7$ for the Pantheon sample and $202=100+6\times17$ for the Pantheon+ sample. The field-average of all the $H_0(\langle\bar{\delta}_\mathrm{m}^I\rangle_k)$ values fitted from $k$-th group reads
\begin{align}
\overline{H}_0\left(\langle\bar{\delta}_\mathrm{m}\rangle_k\equiv\frac{1}{N}\sum_{I=1}^N\langle\bar{\delta}_\mathrm{m}^I\rangle_k\right)\equiv\frac{1}{N}\sum_{I=1}^NH_0(\langle\bar{\delta}_\mathrm{m}^I\rangle_k),
\end{align}
whose correlation to the field-averaged ambient density contrast $\langle\bar{\delta}_\mathrm{m}\rangle_k$ can be linearly fitted with a slope $K$ and intercept $H_0^\mathrm{base}$ by
\begin{align}\label{eq:ObservationalFit}
\bar{\delta}_H\equiv\frac{\overline{H}_0(\langle\bar{\delta}_\mathrm{m}\rangle_k)-H_0^\mathrm{base}}{H_0^\mathrm{base}}=K\langle\bar{\delta}_\mathrm{m}\rangle_k.
\end{align}

We exemplify this linear fitting in the right panel of Fig.~\ref{fig:HostCorrelation} for an illustrative smoothing scale $R=70\,\mathrm{Mpc}/h$. The $H_0(\langle\bar{\delta}_\mathrm{m}^I\rangle_k)$ values with the inclusion (blue) and exclusion (red) of the mass step correction are fitted from the $k$-th group of Pantheon SNe Ia in the $I$-th density field.  The field-averaged $\overline{H}_0(\langle\bar{\delta}_\mathrm{m}\rangle_k)$ values over $N=2000$ density fields are indicated in black with $1\sigma$ error bars $\sigma_{\overline{H}_0}(\langle\bar{\delta}_\mathrm{m}\rangle_k)=\langle\sigma_{H_0}^2(\langle\bar{\delta}_\mathrm{m}^I\rangle_k)\rangle_I^{1/2}$ averaging over all $1\sigma$ uncertainties from all density fields at each density group. Hence, the slope $K$ and intercept $H_0^\mathrm{base}$ can be fitted from these field-averaged data points $(\langle\bar{\delta}_\mathrm{m}\rangle_k,\overline{H}_0(\langle\bar{\delta}_\mathrm{m}\rangle_k),\sigma_{\overline{H}_0}(\langle\bar{\delta}_\mathrm{m}\rangle_k))$. The direct-binned $H_0$ values from all $H_0(\langle\bar{\delta}_\mathrm{m}^I\rangle_k)$ values are also shown in gray. In both cases with and without the mass step correction, there is always a mild linear correlation between the fitted local Hubble constants and the ambient density contrasts of different groups of SNe Ia. The results for the Pantheon+ sample are similar but not presented here for simplicity.

We then trace the scale dependence of this non-local slope in Fig.~\ref{fig:ScaleDependence}. In the left panel, we repeat the previous analysis at $R=15, 20, 30, 40, 50, 60, 70\,\mathrm{Mpc}/h$ for Pantheon(+) samples shown in lighter (darker) colors. Here $R=15\,\mathrm{Mpc}/h$ is roughly the minimal resolution scale of the BOSS density reconstructions. The fitted intercept $H_0^\mathrm{base}$ values are shown with red points, reproducing the usual Hubble tension when compared to the Planck constraint on $H_0$ shown with the red band. More intriguingly, the fitted slopes $K$ (blue points) become more and more positive at larger and larger scale $R$, which is in direct contrast to our theoretical estimation~\eqref{eq:TheoreticalSlope} shown with a green dashed line. The $1\sigma$ error bars in $K(R)$ are extracted from diagonal elements of an inverse covariance matrix characterizing the correlations between different smoothing scales as detailed in the supplemental material~\cite{footnote}. The deviation significance between the observational $K(R)$ and $\Lambda$CDM prediction at each smoothing scale is quantified in the right panel with the largest deviation significance close to $3\sigma$ at $R=60\,\mathrm{Mpc}/h$, though the averaged deviation significance is $1.76\sigma$ ($1.44\sigma$) for Pantheon(+) samples. Therefore, this observational $K(R)$ is still marginally consistent with the $\Lambda$CDM expectation, but future galaxy surveys would enlarge the SN sample and improve the density reconstructions to further reduce the scattering in fitting this non-local slope.

\textbf{Conclusion and discussion.}---
The current Hubble tension has drawn much attention recently for model buildings and systematics checks, among which the calibration errors are claimed to be well controlled, while the physical origin for the scatter in the SN standardization remains mysterious especially for the Hubble residual correlation to the stellar mass of the SN-host galaxy. Since the more massive halos that usually contain the more massive SN-host galaxies tend to populate in the denser regions than the less massive halos, a Hubble residual correlation to the ambient density contrast of SN-host galaxy might be expected. By fitting the local Hubble constant to a group of Pantheon(+) SNe Ia selected with the same ambient density contrast of their SN-host galaxies at a given scale, we have revealed this host-galaxy ambient density correlation with and without the mass step corrections.  We have also found that this residual linear correlation becomes more and more positive at larger and larger scales, on the contrary to the slightly negative correlation predicted by the $\Lambda$CDM model. Several comments are given below concerning about this new Hubble-variation correlation and its scale dependence:

First, although the host density correlation we consider is originally motivated from the host mass correlation, it is there in the data regardless of the inclusion or exclusion of the mass step correction, yielding this host density correlation as a new independent residual correlation from the usual host mass correlation instead of a direct inheritance of the latter one. This is not surprising since it is rather indirect via the host halo mass of the SN-host galaxy to relate the stellar mass of the SN-host galaxy to its ambient density contrast estimated at a given scale.

Second, this Hubble-variation correlation perfectly matches our theoretical estimation at small scales but becomes more and more positive when going to the larger scales, indicating some uncounted corrections to the peculiar velocity of the SN-host galaxy from its ambient density contrast at larger scales, for example, large-scale external flow~\cite{Peterson:2021hel}, and it cannot be of astrophysical origins alone since the scale at which we estimate the ambient density contrast of the SN-host galaxy is much larger than any astrophysical scales.

Third, if there is no new systematics found to contribute to the peculiar velocity of the SN-host galaxy, this new Hubble-variation correlation and its scale dependence could be a smoking gun of new physics. For example, a recently proposed model of the chameleon dark energy~\cite{Cai:2021wgv} admits a higher effective cosmological constant in the denser regions, driving the dense regions to expand locally faster than the less dense regions. Since all regions become less dense when tracing back in time, this model could naturally predict a smaller early-time background Hubble constant than the late-time local measurements.

\begin{acknowledgments}
We sincerely thank Guilhem Lavaux and Jens Jasche for kindly generating the BOSS density reconstruction data for us without both RSD and light cone effects. We also thank David Jones and Dillon Brout for the correspondence on the treatment of the mass step correction in the Pantheon(+) data, and Bin Hu and Qi Guo for his and her stimulating discussions.
This work is supported by the National Key Research and Development Program of China Grant  No. 2021YFC2203004, No. 2020YFC2201501 and No. 2021YFA0718304, 
the National Natural Science Foundation of China Grants No. 12105344, No. 12122513, No. 11991052, and No. 12047503, the Strategic Priority Research Program of the Chinese Academy of Sciences (CAS) Grant No. XDPB15 and the CAS Project for Young Scientists in Basic Research YSBR-006, the Key Research Program of Frontier Sciences of CAS, 
and the Science Research Grants from the China Manned Space Project with No. CMS-CSST-2021-B01.
We acknowledge the use of the High Performance Cluster at ITP-CAS.
\end{acknowledgments}

\appendix

\section{Appendix A. The Hubble-constant variation}\label{app:HubbleVariation}

\begin{figure*}
\centering
\includegraphics[width=0.8\textwidth]{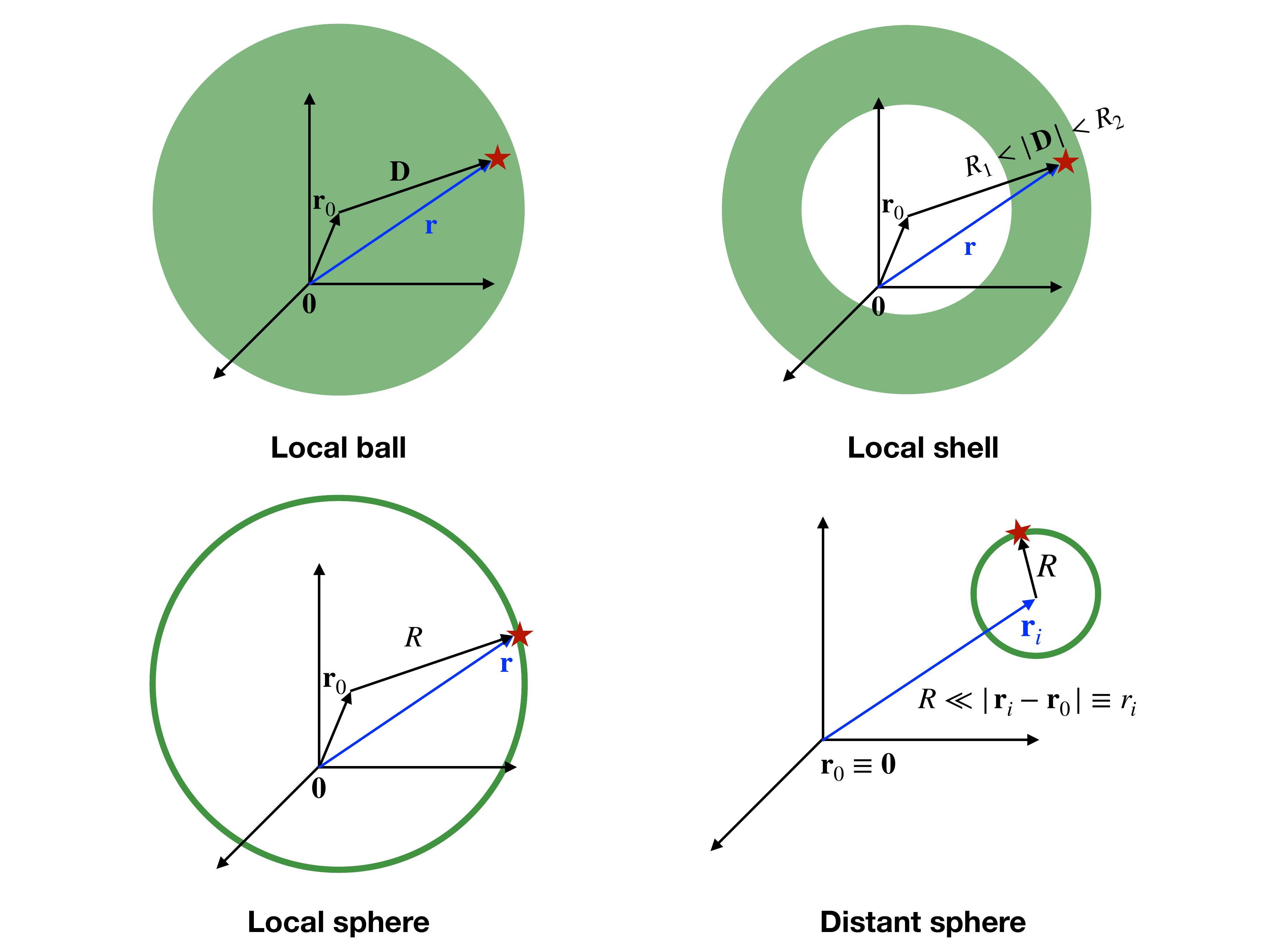}\\
\caption{The schematics for the Hubble-constant variations from continuous samples of local distance indicators within a local ball (top left), a local shell (top right), a local sphere (bottom left), and a distant sphere (bottom right).}\label{fig:LocalContinuous}
\end{figure*}

In this appendix, we will derive for the first time in the $\Lambda$CDM model a theoretical estimation on the variations in the measured local Hubble constants from an arbitrary discrete sample of distant SNe Ia with the same ambient density contrast estimated at a given scale. Before we dive into the details of the theoretical estimation on the Hubble-constant variation from a discrete sample of distant SNe Ia, we first look into the Hubble-constant variations from continuous samples of local distance indicators within a local ball, a local shell, a local sphere, and distant sphere. See Fig.~\ref{fig:LocalContinuous} for a schematics demonstration.

\subsection{The Hubble-constant variations from a local ball and a local point}

To estimate the local Hubble expansion rate $H_0^\mathrm{loc}(\mathbf{r}_0)$ at position $\mathbf{r}_0$, one can first measure the relative redshift $z_i$ and distance $D_i\equiv|\mathbf{D}_i|=|\mathbf{r}_i-\mathbf{r}_0|$ for a sample of local distance indicators at positions $\mathbf{r}_i (i=1,2,\cdots,N)$ within a 3-ball $B_R^3(\mathbf{r}_0)$ of radius $R$, then the local Hubble constant can be approximated  by the mean value~\cite{1992AJ....103.1427T}
\begin{align}
H_0^\mathrm{loc}(\mathbf{r}_0)\approx\overline{H}_0^\mathrm{loc}(\mathbf{r}_0)=H_0^\mathrm{bac}+\frac{1}{N}\sum_{i=1}^N\frac{\mathbf{v}_i\cdot(\mathbf{r}_i-\mathbf{r}_0)}{|\mathbf{r}_i-\mathbf{r}_0|^2},
\end{align}
where $\mathbf{v}_i$ is the peculiar velocity on top of the background  Hubble-flow velocity $H_0^\mathrm{bac}D_i$ satisfying $cz_i=H_0^\mathrm{bac}D_i+\mathbf{v}_i\cdot\hat{\mathbf{D}}_i$. This operational definition would necessarily lead to a local variation in the measured Hubble constant $\bar{\delta}_H(\mathbf{r}_0)\equiv(\overline{H}_0^\mathrm{loc}(\mathbf{r}_0)-H_0^\mathrm{bac})/H_0^\mathrm{bac}$, or in a continuous form as~\cite{Shi:1995nq,Shi:1997aa,Wang:1997tp}
\begin{align}
\bar{\delta}_H(\mathbf{r}_0; B_R^3(\mathbf{r}_0))=\frac{1}{H_0^\mathrm{bac}}\int\mathrm{d}^3\mathbf{r}\frac{\mathbf{v}(\mathbf{r})\cdot(\mathbf{r}-\mathbf{r}_0)}{|\mathbf{r}-\mathbf{r}_0|^2}W_R(\mathbf{r}-\mathbf{r}_0)
\end{align}
for a continuous peculiar velocity field $\mathbf{v}(\mathbf{r})$, where the window function $W_R(\mathbf{D})=(4\pi R^3/3)^{-1}\Theta(R-D)$ with the step function $\Theta(R-D)$ is used for selecting all local distance indicators within a ball of radius $R$.

For the $\Lambda$CDM model, the linear perturbation theory has related the Fourier mode of the peculiar velocity field $\mathbf{v}(\mathbf{r})=(2\pi)^{-3}\int\mathrm{d}^3\mathbf{k}\,\widetilde{\mathbf{v}}(\mathbf{k})e^{i\mathbf{k}\cdot\mathbf{r}}$ to the Fourier mode of the density contrast field $\delta_\mathrm{m}(\mathbf{r})\equiv(\rho_\mathrm{m}(\mathbf{r})-\bar{\rho}_\mathrm{m})/\bar{\rho}_\mathrm{m}=(2\pi)^{-3}\int\mathrm{d}^3\mathbf{k}\,\widetilde{\delta}_\mathrm{m}(\mathbf{k})e^{i\mathbf{k}\cdot\mathbf{r}}$ by the so-called Peebles relation~\cite{1980lssu.book.....P,1993ppc..book.....P}
\begin{align}\label{eq:Peebles}
\widetilde{\mathbf{v}}(\mathbf{k})=if(\Omega_\mathrm{m})H_0^\mathrm{bac}\widetilde{\delta}_\mathrm{m}(\mathbf{k})\mathbf{k}/k^2,
\end{align}
where the growth factor $f(\Omega_\mathrm{m})\approx\Omega_\mathrm{m}^{0.55}$ from the $\Lambda$CDM model will be abbreviated simply as $f$ hereafter. Hence the local Hubble-constant variation can be expressed as
\begin{align}\label{eq:deltaHR}
\bar{\delta}_H(\mathbf{r}_0;B_R^3(\mathbf{r}_0))=f\int\frac{\mathrm{d}^3\mathbf{k}}{(2\pi)^3}\widetilde{\delta}_\mathrm{m}(\mathbf{k})e^{i\mathbf{k}\cdot\mathbf{r}_0}\widetilde{\mathcal{L}}(kR),
\end{align}
where the window function $\widetilde{\mathcal{L}}(kR)$ is evaluated as
\begin{align}
\widetilde{\mathcal{L}}(kR)&=\int\mathrm{d}^3\mathbf{D}\frac{i\mathbf{k}\cdot\mathbf{D}}{k^2D^2}\frac{\Theta(R-D)}{\frac43\pi R^3}e^{i\mathbf{k}\cdot\mathbf{D}},\\
&=\int_0^R\mathrm{d}D\frac{iD/k}{\frac43\pi R^3}\int_0^{2\pi}\mathrm{d}\varphi\int_0^\pi\mathrm{d}\theta\sin\theta\cos\theta e^{ikD\cos\theta},\nonumber\\
&=\frac{3}{(kR)^3}\left(\sin kR-\int_0^{kR}\mathrm{d}x\frac{\sin x}{x}\right)\\
&\equiv\frac{3}{(kR)^3}\left(\sin kR-\mathrm{Si}\,kR\right).
\end{align}
For sufficiently local measurements at $\mathbf{r}_0$ with $R\to0$, the window function approaches to $\widetilde{\mathcal{L}}(kR)\to-1/3$, therefore, the well-known Turner-Cen-Ostriker (TCO) relation~\cite{1992AJ....103.1427T,Shi:1995nq,Shi:1997aa,Wang:1997tp} is thus derived,
\begin{align}\label{eq:deltaHR0}
\bar{\delta}_H(\mathbf{r}_0;B_{R\to0}^3(\mathbf{r}_0))&=-\frac{f}{3}\int\frac{\mathrm{d}^3\mathbf{k}}{(2\pi)^3}\widetilde{\delta}_\mathrm{m}(\mathbf{k})e^{i\mathbf{k}\cdot\mathbf{r}_0}\nonumber\\
&=-\frac{f}{3}\delta_\mathrm{m}(\mathbf{r}_0),
\end{align}
which can also be recovered from going to the real space as we will derive below.

For a finite $R$, we can further evaluate the local Hubble-constant variation \eqref{eq:deltaHR} as
\begin{align}
\bar{\delta}_H(\mathbf{r}_0;B_R^3(\mathbf{r}_0))=f\int\mathrm{d}^3\mathbf{r}\,\delta_\mathrm{m}(\mathbf{r})\int\frac{\mathrm{d}^3\mathbf{k}}{(2\pi)^3}e^{-i\mathbf{k}\cdot(\mathbf{r}-\mathbf{r}_0)}\widetilde{\mathcal{L}}(kR)\nonumber
\end{align}
by inserting back the inverse Fourier transform of the density contrast mode $\widetilde{\delta}_\mathrm{m}(\mathbf{k})=\int\mathrm{d}^3\mathbf{r}\,\delta_\mathrm{m}(\mathbf{r})e^{-i\mathbf{k}\cdot\mathbf{r}}$. After abbreviating $\mathbf{D}=\mathbf{r}-\mathbf{r}_0$, the second integral
\begin{align}
\int\frac{\mathrm{d}^3\mathbf{k}}{(2\pi)^3}e^{-i\mathbf{k}\cdot\mathbf{D}}\widetilde{\mathcal{L}}(kR)&=\int\frac{k^2\mathrm{d}k}{(2\pi)^3}\frac{3}{(kR)^3}(\sin kR-\mathrm{Si}\,kR)\nonumber\\
&\times\int_0^{2\pi}\mathrm{d}\varphi\int_0^\pi\mathrm{d}\theta\sin\theta e^{-ikD\cos\theta}\nonumber
\end{align}
can be performed first by changing to the new variable $x\equiv kR$ with a ratio $\lambda\equiv D/R$ as
\begin{align}
\int\frac{\mathrm{d}^3\mathbf{k}}{(2\pi)^3}e^{-i\mathbf{k}\cdot\mathbf{D}}\widetilde{\mathcal{L}}(kR)&=\frac{3}{2\pi^2R^3}\int_0^\infty\mathrm{d}x\left(\frac{\sin x}{x}-\frac{\mathrm{Si}\,x}{x}\right)\frac{\sin\lambda x}{\lambda x},\nonumber
\end{align}
which is mathematically equivalent to
\begin{align}
\int\frac{\mathrm{d}^3\mathbf{k}}{(2\pi)^3}e^{-i\mathbf{k}\cdot\mathbf{D}}\widetilde{\mathcal{L}}(kR)=\frac{\Theta(1-\lambda)}{\frac43\pi R^3}\ln\lambda\equiv W_R(\mathbf{D})\ln\frac{D}{R}.\nonumber
\end{align}
Therefore, we find that the local Hubble-constant variation \eqref{eq:deltaHR} can be generally obtained in the real space as
\begin{align}\label{eq:deltaHReal}
\bar{\delta}_H(\mathbf{r}_0;B_R^3(\mathbf{r}_0))=f\int\mathrm{d}^3\mathbf{D}\,\delta_\mathrm{m}(\mathbf{r}_0+\mathbf{D})W_R(\mathbf{D})\ln\frac{D}{R},
\end{align}
indicating that using a continuous sample of local distance indicators within a sphere of radius $R$ would necessarily lead to a local variation in the measured Hubble constant proportional to the weighted density contrast  averaged within the radius $R$. 
In particular, for sufficiently local measurements at $\mathbf{r}_0$ in the limit $0<D<R\to0$, the density contrast $\delta_\mathrm{m}(\mathbf{r}_0+\mathbf{D})\to\delta_\mathrm{m}(\mathbf{r}_0)$ can be factorized out, and the resulted local Hubble-constant variation
\begin{align}
\bar{\delta}_H(\mathbf{r}_0;B_{R\to0}^3(\mathbf{r}_0))&=f\delta_\mathrm{m}(\mathbf{r}_0)\int_0^R\frac{4\pi D^2\mathrm{d}D}{\frac43\pi R^3}\ln\frac{D}{R}\nonumber\\
&=-\frac{f}{3}\delta_\mathrm{m}(\mathbf{r}_0)
\end{align}
exactly reproduces the well-known TCO relation~\eqref{eq:deltaHR0}. Unfortunately, for a finite $R$, the local Hubble-constant variation \eqref{eq:deltaHReal} in real space cannot be evaluated further due to the random values taken by the density contrast field $\delta_\mathrm{m}(\mathbf{r}_0+\mathbf{D})$. However, this is not the case for a continuous sample of local distance indicators constrained on a local sphere as we will elaborate below.

\subsection{The Hubble-constant variations from a local shell and a local sphere}

We can generalize the above derivations into the case where the selected local distance indicators are located within a shell $B_{R_2}^3(\mathbf{r}_0)\backslash B_{R_1}^3(\mathbf{r}_0)\equiv\mathrm{Sh}_{R_1}^{R_2}(\mathbf{r}_0)$ with innermost and outermost radii being $R_1$ and $R_2$, respectively, then with the window function
\begin{align}
W(\mathbf{D};R_1,R_2)=\frac{\Theta(R_2-D)\Theta(D-R_1)}{\frac43\pi(R_2^3-R_1^3)},
\end{align}
the corresponding variation in the Hubble constant
\begin{align}
\bar{\delta}_H(\mathbf{r}_0;\mathrm{Sh}_{R_1}^{R_2}(\mathbf{r}_0))=\int\mathrm{d}^3\mathbf{D}\frac{\mathbf{v}(\mathbf{r}_0+\mathbf{D})\cdot\mathbf{D}}{H_0^\mathrm{bac}D^2}W(\mathbf{D};R_1,R_2)
\end{align}
can be similarly evaluated in the momentum space as
\begin{align}
\bar{\delta}_H(\mathbf{r}_0;\mathrm{Sh}_{R_1}^{R_2}(\mathbf{r}_0))=f\int\frac{\mathrm{d}^3\mathbf{k}}{(2\pi)^3}\widetilde{\delta}_\mathrm{m}(\mathbf{k})e^{i\mathbf{k}\cdot\mathbf{r}_0}\widetilde{\mathcal{L}}(\mathbf{k};R_1,R_2)
\end{align}
with the window function of the form
\begin{align}
\widetilde{\mathcal{L}}(\mathbf{k};R_1,R_2)=\frac{3}{k^3(R_2^3-R_1^3)}\left.\left(\sin kR-\mathrm{Si}\,kR\right)\right|_{R_1}^{R_2},
\end{align}
or in the real space as
\begin{align}\label{eq:deltaHshell}
\bar{\delta}_H(\mathbf{r}_0;\mathrm{Sh}_{R_1}^{R_2}(\mathbf{r}_0))=f\int\mathrm{d}^3\mathbf{D}\,\delta_\mathrm{m}(\mathbf{r}_0+\mathbf{D})\mathcal{L}(\mathbf{D};R_1,R_2)
\end{align}
with the window function of the form
\begin{align}
&\mathcal{L}(\mathbf{D};R_1,R_2)=\int\frac{\mathrm{d}^3\mathbf{k}}{(2\pi)^3}e^{-i\mathbf{k}\cdot\mathbf{D}}\widetilde{\mathcal{L}}(\mathbf{k};R_1,R_2)\\
&=W(\mathbf{D};R_1,R_2)\ln\frac{D}{R_2}+\frac{\Theta(R_1-D)}{\frac43\pi(R_2^3-R_1^3)}\ln\frac{R_1}{R_2}.
\end{align}

In particular, for the distance indicators constrained within an infinitely thin shell in the limit $R_1\to R_2\equiv R$, 
\begin{align}
\lim_{R_1\to R_2\equiv R}\widetilde{\mathcal{L}}(\mathbf{k}; R_1, R_2)&=-\frac{j_1(kR)}{kR},\\
\lim_{R_1\to R_2\equiv R}\mathcal{L}(\mathbf{D};R_1,R_2)&=-\frac13W_R(\mathbf{D}),
\end{align}
we can derive the Hubble-constant variation from the distance indicators on a sphere of a radius $R$ at $\mathbf{r}_0$ as
\begin{align}\label{eq:deltaHlocalS2}
\bar{\delta}_H(\mathbf{r}_0;S_R^2(\mathbf{r}_0))=-\frac{f}{3}\bar{\delta}_\mathrm{m}^R(\mathbf{r}_0),
\end{align}
where the ambient density contrast $\bar{\delta}_\mathrm{m}^R(\mathbf{r}_0)$ is the averaged density contrast within a local sphere of radius $R$,
\begin{align}
\bar{\delta}_\mathrm{m}^R(\mathbf{r}_0)=\int\mathrm{d}^3\mathbf{D}\,\delta_\mathrm{m}(\mathbf{r}_0+\mathbf{D})W_R(\mathbf{D}).
\end{align}
This new relation \eqref{eq:deltaHlocalS2} can also be derived directly from
\begin{align}
&\bar{\delta}_H(\mathbf{r}_0;S_R^2(\mathbf{r}_0))
=\int\mathrm{d}^3\mathbf{D}\frac{\mathbf{v}(\mathbf{r}_0+\mathbf{D})\cdot\mathbf{D}}{H_0^\mathrm{bac}D^2}\frac{\delta(D-R)}{4\pi R^2}\nonumber\\
&=\int\mathrm{d}^3\mathbf{D}\int\frac{\mathrm{d}^3\mathbf{k}}{(2\pi)^3}e^{i\mathbf{k}\cdot(\mathbf{r}_0+\mathbf{D})}\frac{\widetilde{\mathbf{v}}(\mathbf{k})\cdot\mathbf{D}}{H_0^\mathrm{bac}D^2}\frac{\delta(D-R)}{4\pi R^2}\nonumber\\
&=\int\frac{\mathrm{d}^3\mathbf{k}}{(2\pi)^3}e^{i\mathbf{k}\cdot\mathbf{r}_0}\frac{if\widetilde{\delta}_\mathrm{m}(\mathbf{k})}{k^2}\int\mathrm{d}^3\mathbf{D}\,e^{i\mathbf{k}\cdot\mathbf{D}}\frac{\mathbf{k}\cdot\mathbf{D}}{D^2}\frac{\delta(D-R)}{4\pi R^2}\nonumber\\
&=f\int\mathrm{d}^3\mathbf{r}\,\delta_\mathrm{m}(\mathbf{r}_0+\mathbf{r})\int\frac{\mathrm{d}^3\mathbf{k}}{(2\pi)^3}e^{-i\mathbf{k}\cdot\mathbf{r}}\frac{i}{k^2}(ik^2)\frac{j_1(kR)}{kR}\nonumber\\
&=f\int\mathrm{d}^3\mathbf{r}\,\delta_\mathrm{m}(\mathbf{r}_0+\mathbf{r})\left(-\frac13W_R(\mathbf{r})\right)
\equiv-\frac{f}{3}\bar{\delta}_\mathrm{m}^R(\mathbf{r}_0)
\end{align}
by replacing the shell window function $W(\mathbf{D};R_1,R_2)$ with a Dirac delta function $\delta(D-R)$ confined on a sphere. In the second line, we have gone to the momentum space of the peculiar velocity field, which has been replaced in the third line by the Fourier mode of the density contrast field according to the Peebles relation~\eqref{eq:Peebles}. In the fourth line, the integration over $\mathbf{D}$ is first performed by choosing $\mathbf{k}$ as the $z$-axis, and then the density contrast field has been put back to the real space around $\mathbf{r}_0$. In the last line, the integration over $\mathbf{k}$ is obtained by choosing $\mathbf{r}$ as the $z$-axis. Note that technique we take for going back and forth between the momentum space and real space will be very useful when dealing with other examples below.

In estimating $\bar{\delta}_\mathrm{m}^R(\mathbf{r}_0)$ from the real data, it would quickly converges to 0 after averaging the density contrast over a larger and larger radius $R$. Therefore, using a distance indicator sample sufficiently distant would largely reduce the measured Hubble-constant variation. This is why the local distance ladder measurements would select SNe Ia on the Hubble flow in the redshift range $0.0233<z<0.15$ with the lower redshift $z=0.0233$ corresponding to the homogeneity scale $R_\mathrm{homo}=70\,\mathrm{Mpc}/h$~\cite{Scrimgeour:2012wt}, where the fractal dimension~\cite{Hogg:2004vw} of the galaxy number counting in such a sphere deviates only one percent from the expectation of the cosmic homogeneity, so that the measured Hubble-constant variation can be reduced below the percent level.

\subsection{The Hubble-constant variation from a distant sphere}

\begin{figure*}
\centering
\includegraphics[width=0.8\textwidth]{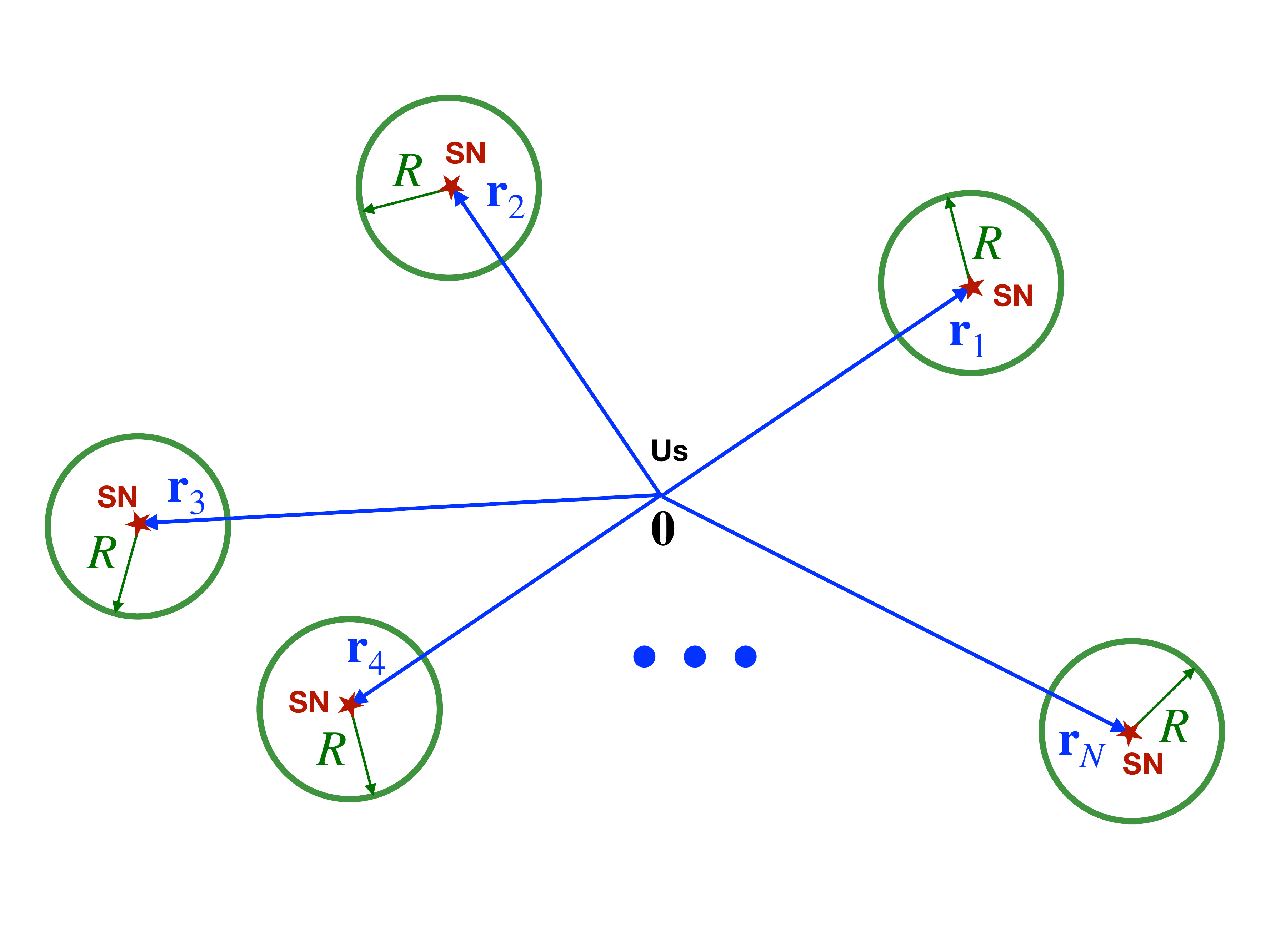}\\
\caption{The schematics for the Hubble-constant variation from a discrete sample of local distance indicators at $\mathbf{r}_i (i=1,2,\cdots,N)$ estimated at a scale $R\ll r_i$.}\label{fig:DistantDiscrete}
\end{figure*}

An immediate generalization of the Hubble-constant variation from a local sphere would be a displaced one,
\begin{align}\label{eq:farS2}
\bar{\delta}_H(\mathbf{r}_0;S^2_R(\mathbf{r}_i))=\int\mathrm{d}^3\mathbf{D}\frac{\mathbf{v}(\mathbf{r}_i+\mathbf{D})\cdot(\mathbf{r}_i+\mathbf{D})}{H_0^\mathrm{bac}|\mathbf{r}_i+\mathbf{D}|^2}\frac{\delta(D-R)}{4\pi R^2},
\end{align}
where the local distance indicators are located on a sphere $S^2_R(\mathbf{r}_i)$ of radius $R$ centered at $\mathbf{r}_i$ different from the observer at $\mathbf{r}_0\equiv\mathbf{0}$.  At the limit of a distant small sphere, $R\ll|\mathbf{r}_i-\mathbf{r}_0|\equiv r_i$ so that $|\mathbf{r}_i+\mathbf{D}|^2\approx r_i^2$, the peculiar velocity field $\mathbf{v}(\mathbf{r}_i+\mathbf{D})$ can be expanded in the neighborhood of $\mathbf{r}_i$ to the first order in $\mathbf{D}$ as 
\begin{align}\label{eq:1stOrder}
\mathbf{v}(\mathbf{r}_i+\mathbf{D})\approx \mathbf{v}(\mathbf{r}_i)+(\mathbf{\nabla}_\mathbf{r}\otimes\mathbf{v}(\mathbf{r}_i))\mathbf{D}, 
\end{align}
where  $(\mathbf{\nabla}_\mathbf{r}\otimes\mathbf{v}(\mathbf{r}_i))^a_{\,\,b}=\partial v^a(\mathbf{r}_i)/\partial r^b$  is the Jacobian matrix given by the partial derivative of $\mathbf{v}\equiv(v_x,v_y,v_z)$ with respect to $\mathbf{r}\equiv(x,y,z)$ at $\mathbf{r}_i$. For the Fourier transformation $\mathbf{v}(\mathbf{r})=(2\pi)^{-3}\int\mathrm{d}^3\mathbf{k}\,\widetilde{\mathbf{v}}(\mathbf{k})e^{i\mathbf{k}\cdot\mathbf{r}}$, the Jacobian matrix term reads
\begin{align}
(\mathbf{\nabla}_\mathbf{r}\otimes\mathbf{v}(\mathbf{r}_i))\mathbf{D}&=\int\frac{\mathrm{d}^3\mathbf{k}}{(2\pi)^3}e^{i\mathbf{k}\cdot\mathbf{r}_i}[i\mathbf{k}\cdot(\mathbf{\nabla}_\mathbf{r}\otimes\mathbf{r}_i)\mathbf{D}]\widetilde{\mathbf{v}}(\mathbf{k})\nonumber\\
&=\int\frac{\mathrm{d}^3\mathbf{k}}{(2\pi)^3}e^{i\mathbf{k}\cdot\mathbf{r}_i}(i\mathbf{k}\cdot\mathbf{D})\widetilde{\mathbf{v}}(\mathbf{k})\label{eq:Jacobian}
\end{align}
since $(\mathbf{\nabla}_\mathbf{r}\otimes\mathbf{r}_i)=\mathrm{diag}(1,1,1)$ is simply an identity matrix. With above approximations in mind, we can split the integral~\eqref{eq:farS2} into four main contributions,
\begin{align}\label{eq:deltaH1stOrder}
&\bar{\delta}_H(\mathbf{0};S_R^2(\mathbf{r}_i))\approx\int\mathrm{d}^3\mathbf{D}\frac{\delta(D-R)}{4\pi R^2}\left\{\frac{\mathbf{v}(\mathbf{r}_i)\cdot\mathbf{r}_i}{H_0^\mathrm{bac}r_i^2}+\frac{\mathbf{v}(\mathbf{r}_i)\cdot\mathbf{D}}{H_0^\mathrm{bac}r_i^2}\right.\nonumber\\
&+\left.\frac{[(\mathbf{\nabla}_\mathbf{r}\otimes\mathbf{v}(\mathbf{r}_i))\mathbf{D}]\cdot\mathbf{r}_i}{H_0^\mathrm{bac}r_i^2}+\frac{[(\mathbf{\nabla}_\mathbf{r}\otimes\mathbf{v}(\mathbf{r}_i))\mathbf{D}]\cdot\mathbf{D}}{H_0^\mathrm{bac}r_i^2}\right\}.
\end{align}

The first contribution is simply the Hubble-constant variation measured from a single distant indicator at $\mathbf{r}_i$,
\begin{align}
\int\mathrm{d}^3\mathbf{D}\frac{\delta(D-R)}{4\pi R^2}\frac{\mathbf{v}(\mathbf{r}_i)\cdot\mathbf{r}_i}{H_0^\mathrm{bac}r_i^2}=\frac{\mathbf{v}(\mathbf{r}_i)\cdot\mathbf{r}_i}{H_0^\mathrm{bac}r_i^2}.
\end{align}
The second contribution can be directly calculated as
\begin{align}
&\int\mathrm{d}^3\mathbf{D}\frac{\mathbf{v}(\mathbf{r}_i)\cdot\mathbf{D}}{H_0^\mathrm{bac}r_i^2}\frac{\delta(D-R)}{4\pi R^2}\nonumber\\
&=\int\mathrm{d}^3\mathbf{D}\int\frac{\mathrm{d}^3\mathbf{k}}{(2\pi)^3}e^{i\mathbf{k}\cdot\mathbf{r}_i}\frac{\widetilde{\mathbf{v}}(\mathbf{k})\cdot\mathbf{D}}{H_0^\mathrm{bac}r_i^2}\frac{\delta(D-R)}{4\pi R^2}\nonumber\\
&=f\int\frac{\mathrm{d}^3\mathbf{k}}{(2\pi)^3}\widetilde{\delta}_\mathrm{m}(\mathbf{k})e^{i\mathbf{k}\cdot\mathbf{r}_i}\int\mathrm{d}^3\mathbf{D}\frac{i\mathbf{k}\cdot\mathbf{D}}{k^2r_i^2}\frac{\delta(D-R)}{4\pi R^2}\nonumber\\
&=0,
\end{align}
where the Peebles relation~\eqref{eq:Peebles} is used, and the integral over $\mathbf{D}$ is simply vanished if we choose $\mathbf{k}$ as the $z$-axis for the integration. The third contribution can also be directly calculated as
\begin{align}
&\int\mathrm{d}^3\mathbf{D}\frac{[(\mathbf{\nabla}_\mathbf{r}\otimes\mathbf{v}(\mathbf{r}_i))\mathbf{D}]\cdot\mathbf{r}_i}{H_0^\mathrm{bac}r_i^2}\frac{\delta(D-R)}{4\pi R^2}\nonumber\\
&=\int\mathrm{d}^3\mathbf{D}\int\frac{\mathrm{d}^3\mathbf{k}}{(2\pi)^3}e^{i\mathbf{k}\cdot\mathbf{r}_i}(i\mathbf{k}\cdot\mathbf{D})\frac{\widetilde{\mathbf{v}}(\mathbf{k})\cdot\mathbf{r}_i}{H_0^\mathrm{bac}r_i^2}\frac{\delta(D-R)}{4\pi R^2}\nonumber\\
&=f\int\frac{\mathrm{d}^3\mathbf{k}}{(2\pi)^3}\widetilde{\delta}_\mathrm{m}(\mathbf{k})e^{i\mathbf{k}\cdot\mathbf{r}_i}\int\mathrm{d}^3\mathbf{D}\frac{(i\mathbf{k}\cdot\mathbf{D})(i\mathbf{k}\cdot\mathbf{r}_i)}{k^2r_i^2}\frac{\delta(D-R)}{4\pi R^2}\nonumber\\
&=0,
\end{align}
where we have used the Jacobian matrix~\eqref{eq:Jacobian} and the Peebles relation~\eqref{eq:Peebles}, and the integral over $\mathbf{D}$ is also vanished if we choose $\mathbf{k}$ as the $z$-axis for the integration. Similarly, the fourth contribution can be calculated as
\begin{align}
&\int\mathrm{d}^3\mathbf{D}\frac{[(\mathbf{\nabla}_\mathbf{r}\otimes\mathbf{v}(\mathbf{r}_i))\mathbf{D}]\cdot\mathbf{D}}{H_0^\mathrm{bac}r_i^2}\frac{\delta(D-R)}{4\pi R^2}\nonumber\\
&=\int\mathrm{d}^3\mathbf{D}\int\frac{\mathrm{d}^3\mathbf{k}}{(2\pi)^3}e^{i\mathbf{k}\cdot\mathbf{r}_i}(i\mathbf{k}\cdot\mathbf{D})\frac{\widetilde{\mathbf{v}}(\mathbf{k})\cdot\mathbf{D}}{H_0^\mathrm{bac}r_i^2}\frac{\delta(D-R)}{4\pi R^2}\nonumber\\
&=-f\int\frac{\mathrm{d}^3\mathbf{k}}{(2\pi)^3}\widetilde{\delta}_\mathrm{m}(\mathbf{k})e^{i\mathbf{k}\cdot\mathbf{r}_i}\int\mathrm{d}^3\mathbf{D}\frac{(\mathbf{k}\cdot\mathbf{D})^2}{k^2r_i^2}\frac{\delta(D-R)}{4\pi R^2}\nonumber\\
&=-f\int\mathrm{d}^3\mathbf{r}\,\delta_\mathrm{m}(\mathbf{r}_i+\mathbf{r})\int\frac{\mathrm{d}^3\mathbf{k}}{(2\pi)^3}e^{-i\mathbf{k}\cdot\mathbf{r}}\left(\frac13\frac{R^2}{r_i^2}\right)\nonumber\\
&=-\frac{f}{3}\delta_\mathrm{m}(\mathbf{r}_i)\frac{R^2}{r_i^2}\simeq-\frac{f}{3}\bar{\delta}_\mathrm{m}^\mathrm{lin}(\mathbf{r}_i)\frac{R^2}{r_i^2},
\end{align}
where the integration over $\mathbf{k}$ simply results in a Dirac delta function $\delta(\mathbf{r})$ that picks out the local density contrast $\delta_\mathrm{m}(\mathbf{r}_i)$ at $\mathbf{r}_i$ in the last line. However, since all the above derivations are carried out in the linear perturbation regime, the would-be Dirac delta function is actually smeared with a small but finite width so that $\delta_\mathrm{m}(\mathbf{r}_i)$ is actually not the local density contrast exactly at $\mathbf{r}_i$ but should be smoothed over the minimal linear scale $R_\mathrm{lin}=8\,\mathrm{Mpc}/h$ around $\mathbf{r}_i$.

Therefore, the local Hubble-constant variation from a continuous sample of local distance indicators on a distant small sphere of radius $R$ centered at $\mathbf{r}_i$ can be estimated analytically as
\begin{align}\label{eq:deltaHfarS2}
\bar{\delta}_H(\mathbf{0};S_R^2(\mathbf{r}_i))\approx\frac{\mathbf{v}(\mathbf{r}_i)\cdot\mathbf{r}_i}{H_0^\mathrm{bac}r_i^2}-\frac{f}{3}\bar{\delta}_\mathrm{m}^\mathrm{lin}(\mathbf{r}_i)\frac{R^2}{r_i^2}.
\end{align}

\subsection{The Hubble-constant variation from a distant discrete sample}

An immediate application of the estimation~\eqref{eq:deltaHfarS2} would be the Hubble-constant variation from a discrete sample of local distance indicators at $\mathbf{r}_i (i=1,2,\cdots,N)$ (see Fig.~\ref{fig:DistantDiscrete}) estimated at a scale $R\ll r_i$ as
\begin{align}
&\bar{\delta}_H(\mathbf{0};\{\mathbf{r}_i|r_i\gg R\})\equiv\frac{1}{N}\sum_{i=1}^N\frac{\mathbf{v}(\mathbf{r}_i)\cdot\mathbf{r}_i}{H_0^\mathrm{bac}r_i^2}\nonumber\\
&\approx\frac{1}{N}\sum_{i=1}^N\left[\bar{\delta}_H(\mathbf{0};S_R^2(\mathbf{r}_i))+\frac{f}{3}\bar{\delta}_\mathrm{m}^\mathrm{lin}(\mathbf{r}_i)\frac{R^2}{r_i^2}\right],
\end{align}
where the first term can be further evaluated as
\begin{align}
\bar{\delta}_H(\mathbf{0};S_R^2(\mathbf{r}_i))
&\approx \int\mathrm{d}^3\mathbf{D}\frac{\mathbf{v}(\mathbf{r}_i+\mathbf{D})\cdot(\mathbf{r}_i+\mathbf{D})}{H_0^\mathrm{bac}r_i^2}\frac{\delta(D-R)}{4\pi R^2}\nonumber\\
&\approx\int\mathrm{d}^3\mathbf{D}\frac{\mathbf{v}(\mathbf{r}_i+\mathbf{D})\cdot\mathbf{D}}{H_0^\mathrm{bac}r_i^2}\frac{\delta(D-R)}{4\pi R^2}\nonumber\\
&=-\frac{f}{3}\bar{\delta}_\mathrm{m}^R(\mathbf{r}_i)\frac{R^2}{r_i^2}.
\end{align}
Here the first approximate equality is taken due to the approximation $|\mathbf{r}_i+\mathbf{D}|\approx r_i^2$ made for $D\to R\ll r_i$. The second line is arrived since the difference between first two lines is approximately vanished,
\begin{align}
&\int\mathrm{d}^3\mathbf{D}\frac{\mathbf{v}(\mathbf{r}_i+\mathbf{D})\cdot\mathbf{r}_i}{H_0^\mathrm{bac}r_i^2}\frac{\delta(D-R)}{4\pi R^2}\nonumber\\
&=\int\mathrm{d}^3\mathbf{D}\int\frac{\mathrm{d}^3\mathbf{k}}{(2\pi)^3}e^{i\mathbf{k}\cdot(\mathbf{r}_i+\mathbf{D})}\frac{\widetilde{\mathbf{v}}(\mathbf{k})\cdot\mathbf{r}_i}{H_0^\mathrm{bac}r_i^2}\frac{\delta(D-R)}{4\pi R^2}\nonumber\\
&=f\int\frac{\mathrm{d}^3\mathbf{k}}{(2\pi)^3}\widetilde{\delta}_\mathrm{m}(\mathbf{k})e^{i\mathbf{k}\cdot\mathbf{r}_i}\frac{i\mathbf{k}\cdot\mathbf{r}_i}{k^2r_i^2}\int\mathrm{d}^3\mathbf{D}\,e^{i\mathbf{k}\cdot\mathbf{D}}\frac{\delta(D-R)}{4\pi R^2}\nonumber\\
&=f\int\mathrm{d}^3\,\mathbf{r}\,\delta_\mathrm{m}(\mathbf{r}_i+\mathbf{r})\int\frac{\mathrm{d}^3\mathbf{k}}{(2\pi)^3}e^{-i\mathbf{k}\cdot\mathbf{r}}\frac{i\mathbf{k}\cdot\mathbf{r}_i}{k^2r_i^2}\frac{\sin kR}{kR}\nonumber\\
&=f\int\mathrm{d}^3\mathbf{r}\,\delta_\mathrm{m}(\mathbf{r}_i+\mathbf{r})\frac{\cos\theta_{\mathbf{r}\mathbf{r}_i}}{4\pi r^2r_i}\Theta(r-R)\approx0,
\end{align}
where we first choose $\mathbf{k}$ as the $z$-axis for the integration over $\mathbf{D}$ and then choose $\mathbf{r}$ as the $z$-axis for the integration over $\mathbf{k}$. We have also used $\cos\theta_{\mathbf{k}\mathbf{r}_i}=\cos\theta_{\mathbf{k}\mathbf{r}}\cos\theta_{\mathbf{r}_i\mathbf{r}}+\sin\theta_{\mathbf{k}\mathbf{r}}\sin\theta_{\mathbf{r}_i\mathbf{r}}\cos(\varphi_{\mathbf{k}\mathbf{r}}-\varphi_{\mathbf{r}_i\mathbf{r}})$ for the $\mathbf{k}\cdot\mathbf{r}_i$ term in the integration over $\mathbf{k}$, where $\theta_{\mathbf{a}\mathbf{b}}$ and $\varphi_{\mathbf{a}\mathbf{b}}$ are the differences in the polar and azimuthal angles between $\mathbf{a}$ and $\mathbf{b}$, respectively.  The final integration is approximately vanished due to the asymmetry of the $\cos\theta_{\mathbf{r}\mathbf{r}_i}$ term and isotropy of $\delta_\mathrm{m}(\mathbf{r}_i+\mathbf{r})$ at large scale $r>R$ under reflection $\mathbf{r}\to-\mathbf{r}$.
The final integral can be directly calculated as
\begin{align}
&\int\mathrm{d}^3\mathbf{D}\frac{\mathbf{v}(\mathbf{r}_i+\mathbf{D})\cdot\mathbf{D}}{H_0^\mathrm{bac}r_i^2}\frac{\delta(D-R)}{4\pi R^2}\nonumber\\
&=\int\mathrm{d}^3\mathbf{D}\int\frac{\mathrm{d}^3\mathbf{k}}{(2\pi)^3}e^{i\mathbf{k}\cdot(\mathbf{r}_i+\mathbf{D})}\frac{\widetilde{\mathbf{v}}(\mathbf{k})\cdot\mathbf{D}}{H_0^\mathrm{bac}r_i^2}\frac{\delta(D-R)}{4\pi R^2}\nonumber\\
&=f\int\frac{\mathrm{d}^3\mathbf{k}}{(2\pi)^3}\widetilde{\delta}_\mathrm{m}(\mathbf{k})e^{i\mathbf{k}\cdot\mathbf{r}_i}\int\mathrm{d}^3\mathbf{D}\frac{i\mathbf{k}\cdot\mathbf{D}}{k^2r_i^2}e^{i\mathbf{k}\cdot\mathbf{D}}\frac{\delta(D-R)}{4\pi R^2}\nonumber\\
&=f\int\mathrm{d}^3\mathbf{r}\,\delta_\mathrm{m}(\mathbf{r}_i+\mathbf{r})\int\frac{\mathrm{d}^3\mathbf{k}}{(2\pi)^3}e^{-i\mathbf{k}\cdot\mathbf{r}}\left[-\frac{j_1(kR)}{kR}\frac{R^2}{r_i^2}\right]\nonumber\\
&=-\frac{f}{3}\int\mathrm{d}^3\mathbf{r}\,\delta_\mathrm{m}(\mathbf{r}_i+\mathbf{r})W_R(\mathbf{r})\frac{R^2}{r_i^2}\nonumber\\
&\equiv-\frac{f}{3}\bar{\delta}_\mathrm{m}^R(\mathbf{r}_i)\frac{R^2}{r_i^2}.
\end{align}
Therefore, the Hubble-constant variation from a discrete sample of local distance indicators at $\mathbf{r}_i (i=1,2,\cdots,N)$ can be estimated analytically at a scale $R\ll r_i$ as
\begin{align}\label{eq:deltaHdiscrete}
\bar{\delta}_H(\mathbf{0};\{\mathbf{r}_i|r_i\gg R\})&\approx-\frac{f}{3}\left\langle\left[\bar{\delta}_\mathrm{m}^R(\mathbf{r}_i)-\bar{\delta}_\mathrm{m}^\mathrm{lin}(\mathbf{r}_i)\right]\frac{R^2}{r_i^2}\right\rangle_i
\end{align}
with the bra-ket symbol denoting the spatial average over all local distance indicators in the given discrete sample.

\section{Appendix B. The slope for the Hubble variation}

We now turn to the observational perspective for testing the theoretical predictions from the $\Lambda$CDM  model on the Hubble-constant variations under different circumstances as we have derived in~\eqref{eq:deltaHReal}, \eqref{eq:deltaHshell}, \eqref{eq:deltaHlocalS2}, \eqref{eq:deltaHfarS2}, and \eqref{eq:deltaHdiscrete} in addition to the well-known TCO relation~\eqref{eq:deltaHR0}, which will be discussed separately below and briefly summarized in Table~\ref{tab:comparison}.

\begin{table*}
\small
\centering
\caption{A brief summary for the cosmic and sample variances, the ability to select a sample, and the prediction on the slope for the  Hubble-constant variation measured by the SN samples from a local position, a local ball, a local shell, a local sphere, a distant sphere, and an arbitrary discrete sample of distant SNe Ia. }
\begin{tabular}{c||c|c|c|c}
\hline
\hline
Sample region & Cosmic variance & Sample variance & Sample selection & Prediction on the slope \\
\hline
\hline
Local position & Large but can be made small & Large & Difficult & Definite\\
\hline
Local ball & Large but can be made small & Large & Easy & Random\\
\hline
Local shell & Large but can be made small & Can be made small & Easy & Random\\
\hline
Local sphere & Large but can be made small & Can be made small & Difficult & Definite\\
\hline
Distant sphere & Small & Can be made small & Difficult & Approximately definite\\
\hline
Distant discrete sample & Small & Small & Easy & Approximately definite\\
\hline
\hline
\end{tabular}
\label{tab:comparison}
\end{table*}

First, although the Hubble-constant variation from our local position~\eqref{eq:deltaHR0} can be theoretically predicted as a definite form $\bar{\delta}_H(\mathbf{r}_0;B_{R\to0}^3(\mathbf{r}_0))=-(f/3)\delta_\mathrm{m}(\mathbf{r}_0)$, it is actually hard to be tested with the real observational data due to (1) a large cosmic variance in $\delta_\mathrm{m}(\mathbf{r}_0)$ induced by the particular choice of a local position $\mathbf{r}_0$, and (2) a large sample variance in $\bar{\delta}_H(\mathbf{r}_0;B_{R}^3(\mathbf{r}_0))$ induced by a small sample volume at $\mathbf{r}_0$ required to be sufficiently local with $R\to0$. The first drawback from the large cosmic variance can be made small by approximating $\bar{\delta}_H(\mathbf{r}_0; B_R^3(\mathbf{r}_0))$ as the averaged $\bar{\delta}_H(\mathbf{r}_i; B_R^3(\mathbf{r}_i))$ over all positions of SNe Ia at $\mathbf{r}_i$. However, the second drawback cannot be made small since the $R\to0$ limit in the TCO relation would necessarily lead to a large sample variance. Nevertheless, it is still widely used in the literature as a practical approximation to test the large local void scenario as a solution to the Hubble tension.

Second, the derived Hubble-constant variations from our local ball~\eqref{eq:deltaHReal} and local shell~\eqref{eq:deltaHshell} also suffer from a large cosmic variance from our local position. The sample variance for the Hubble-constant variation from a local shell can be made small as long as the inner-most radius is larger than the homogeneity scale. However, this is not the case for the Hubble-constant variation from a local ball as it always contains a local point in the center.  Furthermore, no definite theoretical prediction can be made for both cases due to the spatial integration of some weighted density contrast field that is in nature random. For example, the slope defined by the ratio of the Hubble-constant variation with respect to our ambient density contrast,
\begin{align}
K&=f\frac{\int\mathrm{d}^3\mathbf{D}\,\delta_\mathrm{m}(\mathbf{r}_0+\mathbf{D})W_R(\mathbf{D})\ln D/R}{\int\mathrm{d}^3\mathbf{D}\,\delta_\mathrm{m}(\mathbf{r}_0+\mathbf{D})W_R(\mathbf{D})},
\end{align}
cannot conclude whether there is such a correlation from the theoretical ground since the exact value taken by the density contrast field at any particular point is unknown. Nevertheless, we can still define a statistical slope $K$ as
\begin{align}
\frac{\langle K\rangle}{f}
=\frac{\int\mathrm{d}^3\mathbf{D}\int\mathrm{d}\delta_\mathrm{m}\,\delta_\mathrm{m}p(\delta_\mathrm{m},R)W_R(\mathbf{D})\ln D/R}{\int\mathrm{d}^3\mathbf{D}\int\mathrm{d}\delta_\mathrm{m}\,\delta_\mathrm{m}p(\delta_\mathrm{m},R)W_R(\mathbf{D})}
\end{align}
if the probability $p(\delta_\mathrm{m},R)$ for the density contrast field $\delta_\mathrm{m}(\mathbf{r}_0+\mathbf{D})$ to take a given value $\delta_\mathrm{m}$ within $R$ is known. For example, at linear scales, the density contrast field can be described by a Gaussian random field distribution
\begin{align}
p(\delta_\mathrm{m},R)=\frac{1}{\sqrt{2\pi\sigma_R^2}}e^{-\frac{\delta_\mathrm{m}^2}{2\sigma_R^2}}
\end{align}
with its variance $\sigma_R^2$ given by
\begin{align}
\sigma_R^2=\int\frac{k^2\mathrm{d}k}{2\pi^2}P_\mathrm{m}(k)\left[\frac{3j_1(kR)}{kR}\right]^2
\end{align}
from the matter power spectrum $P_\mathrm{m}\mathrm(k)$ at the linear order. As a simple reflection of the cosmological Copernican principle that $\mathbf{r}_0+\mathbf{D}$ is not any special position, the probability $p(\delta_\mathrm{m},R)$ admits no dependence on the specific position $\mathbf{r}_0+\mathbf{D}$ at all, therefore, the integration over $\delta_\mathrm{m}$ can be canceled out in the numerator and denominator, and this statistical slope $\langle K\rangle$ for the Hubble-constant variation from our local ball with respect to our ambient density contrast could have a definite prediction as
\begin{align}
\langle K\rangle=-\frac{f}{3}.
\end{align}
We point out that it is this statistical slope that is actually tested in the literature for the large local void scenario when estimating the Hubble-constant variation from a local sample volume that is not too small.

Third, although the Hubble-constant variation~\eqref{eq:deltaHlocalS2} from a local sphere is exact with a definite form $\bar{\delta}_H(\mathbf{r}_0;S_R^2(\mathbf{r}_0))=-(f/3)\bar{\delta}_\mathrm{m}^R(\mathbf{r}_0)$, it is still generally difficult to be  test directly from our current observational data since the number of the well-observed SNe Ia on the same local sphere centered at our position $\mathbf{r}_0=0$ is usually less than a few, which is too small to yield a sensible constraint on the corresponding Hubble-constant variation. Furthermore, even if we have enough well-observed SNe Ia on our local spheres so that $\bar{\delta}_H(\mathbf{r}_0;S_R^2(\mathbf{r}_0))$ is well-determined for different radii $R$,  it is still not possible to directly infer whether there is such a correlation between the measured Hubble-constant variation and our own ambient density contrast since we only have one local Universe to observe, that is, for any given $R$, there is only one data point $(\bar{\delta}_\mathrm{m}^R(\mathbf{r}_0),\bar{\delta}_H(\mathbf{r}_0;S_R^2(\mathbf{r}_0)))$ to work with, which is not enough to determine whether there is such a slope or not at this scale.

Finally, we propose to test the more general Hubble-constant variation~\eqref{eq:deltaHdiscrete} from a discrete sample of distant SNe Ia but grouped by their own ambient density contrasts at some scales. This proposal could remedy all the aforementioned difficulties: it is not only detached to the specific size and shape of the sample volume, but also free from the large cosmic/sample variances in order to extract the slope of the Hubble-constant variation with respect to their own ambient density contrasts.

For a discrete sample of SNe Ia at the positions $\{\mathbf{r}_i\}  (i=1,2,\cdots,N)$ within a spherical shell $R_1<r_i<R_2$ centered at our position $\mathbf{r}_0\equiv\mathbf{0}$, if the innermost radius $R_1>R_\mathrm{homo}$ is larger than the cosmic homogeneity scale $R_\mathrm{homo}=70\,\mathrm{Mpc}/h$, then the measured Hubble-constant variation from these SNe Ia would be less than one percent as expected analytically from~\eqref{eq:deltaHlocalS2}. However, as we will show shortly below, there is actually a hidden trend (namely a slope) even in this sub-percent Hubble-constant variation if these SNe Ia are selected in such a way that their own ambient density contrasts at a scale $R$ (focusing on the case with $R_\mathrm{lin}<R<R_\mathrm{homo}$) are all equal to the same value $\delta_\mathrm{m}^R$. In this case, the Hubble-constant variation measured from these selected SNe Ia with the same ambient density contrast at a given scale $R$ can be estimated by~\eqref{eq:deltaHdiscrete} as
\begin{align}\label{eq:deltaHdeltam}
\bar{\delta}_H(\mathbf{0};\{\mathbf{r}_i|\bar{\delta}_\mathrm{m}^R(\mathbf{r}_i)=\delta_\mathrm{m}^R\})\approx-\frac{f}{3}\left\langle\frac{R^2}{r_i^2}\right\rangle_i\delta_\mathrm{m}^R,
\end{align}
where the local term $\langle\bar{\delta}_\mathrm{m}^\mathrm{lin}(\mathbf{r}_i)R^2/r_i^2\rangle_i$ has been omitted by checking it numerically in the real data to be smaller than the right-hand-side of~\eqref{eq:deltaHdeltam} since the selection condition  $\bar{\delta}_\mathrm{m}^R(\mathbf{r}_i)=\delta_\mathrm{m}^R$ puts no constraint on the local density contrast $\bar{\delta}_\mathrm{m}^\mathrm{lin}$ at the minimal linear scale. Hence, $\bar{\delta}_\mathrm{m}^\mathrm{lin}$ can be still described by a Gaussian random field distribution with zero mean. This is a good approximation since the BOSS density reconstruction we used for estimating the ambient density contrast of the SN-host galaxy admits a resolution $15\,\mathrm{Mpc}/h$ larger than the minimal linear scale. Therefore, the $\Lambda$CDM model predicts a negative slope
\begin{align}
K&\equiv\frac{\bar{\delta}_H(\mathbf{0};\{\mathbf{r}_i|\bar{\delta}_\mathrm{m}^R(\mathbf{r}_i)=\delta_\mathrm{m}^R\})}{\delta_\mathrm{m}^R}=-\frac{f}{3}\left\langle\frac{R^2}{r_i^2}\right\rangle_i\equiv-\frac{f}{3}Q,
\end{align}
where the magnitude for the mean value of the inverse distance-square can be estimated with a continuous form within a shell as
\begin{align}
Q\equiv\left\langle\frac{R^2}{r_i^2}\right\rangle_i\simeq\int\mathrm{d}^3\mathbf{r}\frac{R^2}{r^2}W(\mathbf{r}; R_1, R_2)=\frac{3(R_2-R_1)R^2}{R_2^3-R_1^3}.
\end{align}
Note here that the slope $K$ can be extracted without a large cosmic variance since now we can have more than one data point to fit the slope $K$ at a given scale $R$ by selecting different samples of SNe Ia with different ambient density contrasts at the same scale $R$.  Changing the scale $R$, we can even trace the evolution of the slope $K$ with respect to the scale $R$. It turns out as a surprise that, using the real data from the selected Pantheon(+) SNe Ia samples grouped by their ambient density contrasts estimated at a given scale from the BOSS density reconstructions, the slope for the measured Hubble-constant variations becomes more and more positively correlated with the ambient density contrasts of the SN-host galaxies at larger and larger scales as we have found in the main context.

\section{Appendix C. Statistical significance}

\begin{figure*}
\centering
\includegraphics[width=0.47\textwidth]{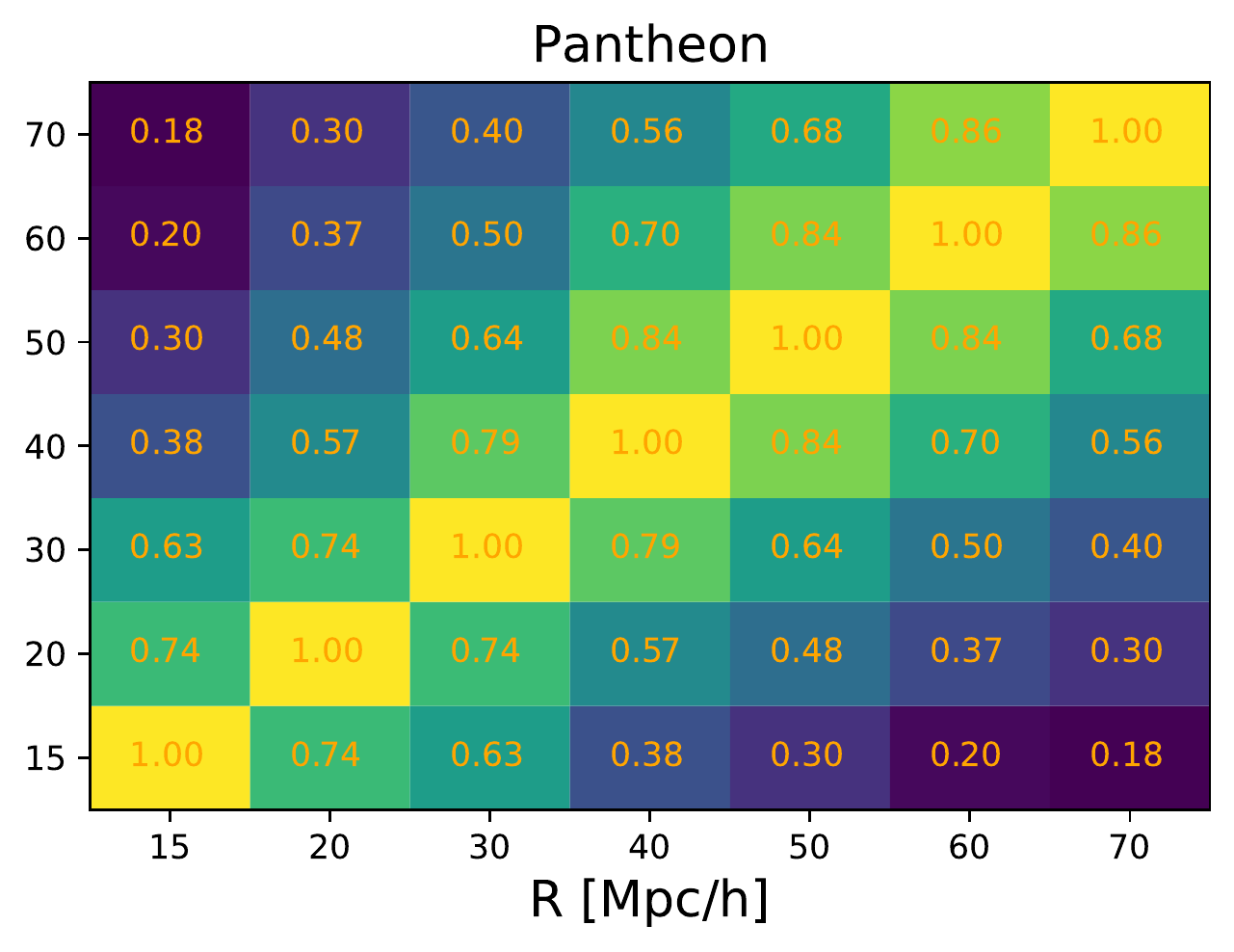}
\includegraphics[width=0.47\textwidth]{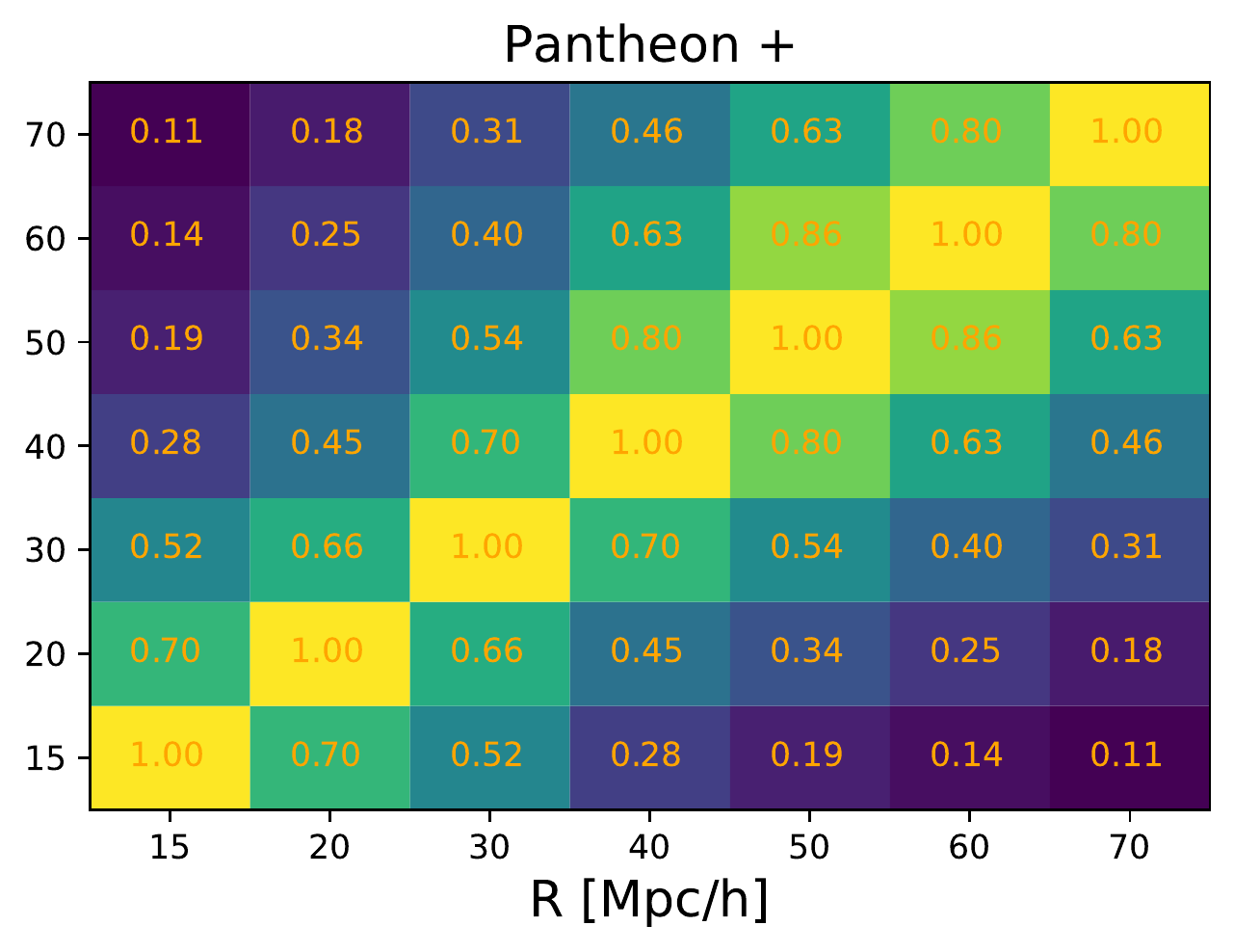}\\
\caption{The correlation matrix of the residual linear slope $K(R)$ between different smoothing scales for the Pantheon (left) and Pantheon+ (right) samples. } \label{fig:correlations}
\end{figure*}

In the main context, with observation data from a total number $M$ of SNe within the total number $N$ of reconstructed density contrast fields,  we have first estimated the ambient density contrast $\bar{\delta}_\mathrm{m}^I(\mathbf{d}_i)$ of each SN at $\mathbf{d}_i$ with a smoothing scale $R$ in the $I$-th density field by averaging over a total number $n_i$ of density contrast $\delta_\mathrm{m}^I(\mathbf{r}_j)$ at $\mathbf{r}_j$ within a sphere of radius $R$ centered at $\mathbf{d}_i$ as
\begin{align}
\bar{\delta}_\mathrm{m}^I(\mathbf{d}_i)=\frac{1}{n_i}\sum_{|\mathbf{r}_j(\mathbf{d}_i)-\mathbf{d}_i|<R,\, j=1}^{n_i}\delta_\mathrm{m}^I(\mathbf{r}_j(\mathbf{d}_i)),
\end{align}
and then presorted these number $M$ of SNe by their ambient density contrast at scale $R$ in the $I$-th density field as $\bar{\delta}_\mathrm{m}^I(\mathbf{d}_{P_1^I})\leq\bar{\delta}_\mathrm{m}^I(\mathbf{d}_{P_2^I})\leq\cdots\leq\bar{\delta}_\mathrm{m}^I(\mathbf{d}_{P_M^I})$, from which we can pick every 100 SNe out as a group each time for fitting the $H_0$ value $H_0(\langle\bar{\delta}_\mathrm{m}^I\rangle_k)$ with $1\sigma$ uncertainty $\sigma_{H_0}(\langle\bar{\delta}_\mathrm{m}^I\rangle_k)$. Here the $k$-th group-averaged ambient density contrast is defined by
\begin{align}
\langle\bar{\delta}_\mathrm{m}^I\rangle_k=\frac{1}{100}\sum_{j=1}^{100}\bar{\delta}_\mathrm{m}^I(\mathbf{d}_{P_{j+(k-1)s}^I})
\end{align}
with neighboring groups shifted by a number $s$ of SNe according to the grouping strategy $M=100+(m-1)s$. Next, we can average the $H_0(\langle\bar{\delta}_\mathrm{m}^I\rangle_k)$ values over all reconstructed density field for the $k$-th group as
\begin{align}
\overline{H}_0\left(\langle\bar{\delta}_\mathrm{m}\rangle_k\equiv\frac{1}{N}\sum_{I=1}^N\langle\bar{\delta}_\mathrm{m}^I\rangle_k\right)\equiv\frac{1}{N}\sum_{I=1}^NH_0(\langle\bar{\delta}_\mathrm{m}^I\rangle_k)
\end{align}
with the corresponding uncertainty of form
\begin{align}
\sigma_{\overline{H}_0}(\langle\bar{\delta}_\mathrm{m}\rangle_k)=\left(\frac{1}{N}\sum_{I=1}^N\sigma_{H_0}^2(\langle\bar{\delta}_\mathrm{m}^I\rangle_k)\right)^{1/2}.
\end{align}
Finally, we can fit these field-averaged data points
\begin{align}
(\langle\bar{\delta}_\mathrm{m}\rangle_k, \overline{H}_0(\langle\bar{\delta}_\mathrm{m}\rangle_k), \sigma_{\overline{H}_0}(\langle\bar{\delta}_\mathrm{m}\rangle_k))
\end{align}
from different density groups $k=1,2,\cdots,m\equiv\frac{M-100}{s}$ by a linear ansatz
\begin{align}\label{eq:linearfit}
\overline{H}_0(\langle\bar{\delta}_\mathrm{m}\rangle_k)=H_0^\mathrm{base}K(R)\langle\bar{\delta}_\mathrm{m}\rangle_k+H_0^\mathrm{base}
\end{align}
with a slope $K(R)$, an intercept $H_0^\mathrm{base}$, and a corresponding $1\sigma$ uncertainty $\sigma_K(R)$. As a comparison, the $\Lambda$CDM prediction for this non-local residual correlation reads
\begin{align}
K_\mathrm{\Lambda CDM}(R)=-\frac{f}{3}Q(R)\equiv-\frac{f}{3}\left(\frac{1}{M}\sum_{i=1}^M\frac{R^2}{d_i^2}\right).
\end{align}

One might naively estimate the deviation significance between the observational $K(R)$ and theoretical $K_{\Lambda\mathrm{CDM}}(R)$ at each smoothing scale $R$ by
\begin{align}
\sigma(R)\equiv\frac{|K(R)-K_{\Lambda\mathrm{CDM}}(R)|}{\sigma_K(R)}.
\end{align}
However, the deviation significance defined in this way could be underestimated as the non-local slope $K(R)$ at different smoothing scales might be correlated with each other if there is an alternative model other than the $\Lambda$CDM model predicting more positive $K(R)$ at larger scale $R$. To quantify this correlation between different smoothing scales, we evaluate the covariance matrix by
\begin{align}
C_{ab}=\frac{1}{N-1}\sum_{I=1}^{N}\left[K_I(R_a)-\overline{K}(R_a)\right]\left[K_I(R_b)-\overline{K}(R_b)\right],
\end{align}
where the non-local residual slope $K_I(R)$ from the $I$-th density field is obtained by fitting the data points $(\langle\bar{\delta}_\mathrm{m}^I\rangle_k, H_0(\langle\bar{\delta}_\mathrm{m}^I\rangle_k), \sigma_{H_0}(\langle\bar{\delta}_\mathrm{m}^I\rangle_k)$, $k=1,2,\cdots,m$ with a linear model $H_0(\langle\bar{\delta}_\mathrm{m}^I\rangle_k)=H_0^{\mathrm{base},I}K_I(R)\langle\bar{\delta}_\mathrm{m}^I\rangle_k+H_0^{\mathrm{base},I}$, and $\overline{K}(R)=(1/N)\sum_I K_I(R)$ is the corresponding mean value over all reconstructed density fields. In Fig.~\ref{fig:correlations}, we present the associated correlation matrix $\mathrm{Corr}_{ab}=C_{ab}/(\sigma_a\sigma_b), \sigma_c=\sqrt{C_{cc}}$ of this non-local slope between different smoothing scales. Therefore, from the unbiased inverse covariance matrix~\cite{Hartlap:2006kj} $\Sigma_{ab}=\frac{N-1}{N-m-2}C_{ab}$, the $1\sigma$ uncertainty of $K(R)$ can be extracted as
\begin{align}
\sigma_\mathrm{Cov}(R_c)=(\Sigma^{-1})_{cc}^{-1/2},
\end{align}
from which the deviation significance for the observational non-local slope $K(R)$ with respect to the $\Lambda$CDM prediction can be estimated at each smoothing scale by
\begin{align}
\sigma(R)\equiv\frac{|K(R)-K_{\Lambda\mathrm{CDM}}(R)|}{\sigma_\mathrm{Cov}(R)}.
\end{align}
Finally, the overall deviation significance between observational $K(R)$ and theoretical $K(R)$ can be estimated as $\sigma=\sqrt{\langle\sigma^2(R)\rangle_R}$ from averaging each deviation significance-square $\sigma(R)^2$ over all smoothing scales, which is $1.76\sigma$ ($1.44\sigma$) for Pantheon(+) samples.

\bibliographystyle{utphys}
\bibliography{ref}

\providecommand{\href}[2]{#2}\begingroup\raggedright\begin{thebibliography}{10}

\bibitem{Moresco:2022phi}
M.~Moresco {\em et~al.}, ``{Unveiling the Universe with Emerging Cosmological
  Probes},'' {\em submitted to Living Reviews in Relativity} (1, 2022) ,
  \href{http://arxiv.org/abs/2201.07241}{{\ttfamily arXiv:2201.07241
  [astro-ph.CO]}}.

\bibitem{Planck:2018vyg}
{\bfseries Planck} Collaboration, N.~Aghanim {\em et~al.}, ``{Planck 2018
  results. VI. Cosmological parameters},''
  \href{http://dx.doi.org/10.1051/0004-6361/201833910}{{\em Astron. Astrophys.}
  {\bfseries 641} (2020) A6}, \href{http://arxiv.org/abs/1807.06209}{{\ttfamily
  arXiv:1807.06209 [astro-ph.CO]}}. [Erratum: Astron.Astrophys. 652, C4
  (2021)].

\bibitem{Riess:2016jrr}
A.~G. Riess {\em et~al.}, ``{A 2.4\% Determination of the Local Value of the
  Hubble Constant},'' \href{http://dx.doi.org/10.3847/0004-637X/826/1/56}{{\em
  Astrophys. J.} {\bfseries 826} no.~1, (2016) 56},
\href{http://arxiv.org/abs/1604.01424}{{\ttfamily arXiv:1604.01424
  [astro-ph.CO]}}.
%%CITATION = ARXIV:1604.01424;%%.

\bibitem{Riess:2018byc}
A.~G. Riess {\em et~al.}, ``{Milky Way Cepheid Standards for Measuring Cosmic
  Distances and Application to Gaia DR2: Implications for the Hubble
  Constant},'' \href{http://dx.doi.org/10.3847/1538-4357/aac82e}{{\em
  Astrophys. J.} {\bfseries 861} no.~2, (2018) 126},
\href{http://arxiv.org/abs/1804.10655}{{\ttfamily arXiv:1804.10655
  [astro-ph.CO]}}.
%%CITATION = ARXIV:1804.10655;%%.

\bibitem{Riess:2018uxu}
A.~G. Riess {\em et~al.}, ``{New Parallaxes of Galactic Cepheids from Spatially
  Scanning the Hubble Space Telescope: Implications for the Hubble Constant},''
  \href{http://dx.doi.org/10.3847/1538-4357/aaadb7}{{\em Astrophys. J.}
  {\bfseries 855} no.~2, (2018) 136},
\href{http://arxiv.org/abs/1801.01120}{{\ttfamily arXiv:1801.01120
  [astro-ph.SR]}}.
%%CITATION = ARXIV:1801.01120;%%.

\bibitem{Riess:2019cxk}
A.~G. Riess, S.~Casertano, W.~Yuan, L.~M. Macri, and D.~Scolnic, ``{Large
  Magellanic Cloud Cepheid Standards Provide a 1\% Foundation for the
  Determination of the Hubble Constant and Stronger Evidence for Physics Beyond
  LambdaCDM},'' \href{http://dx.doi.org/10.3847/1538-4357/ab1422}{{\em
  Astrophys. J.} {\bfseries 876} no.~1, (2019) 85},
\href{http://arxiv.org/abs/1903.07603}{{\ttfamily arXiv:1903.07603
  [astro-ph.CO]}}.
%%CITATION = ARXIV:1903.07603;%%.

\bibitem{Riess:2020fzl}
A.~G. Riess, S.~Casertano, W.~Yuan, J.~B. Bowers, L.~Macri, J.~C. Zinn, and
  D.~Scolnic, ``{Cosmic Distances Calibrated to 1\% Precision with Gaia EDR3
  Parallaxes and Hubble Space Telescope Photometry of 75 Milky Way Cepheids
  Confirm Tension with $\Lambda$CDM},''
  \href{http://dx.doi.org/10.3847/2041-8213/abdbaf}{{\em Astrophys. J. Lett.}
  {\bfseries 908} no.~1, (2021) L6},
  \href{http://arxiv.org/abs/2012.08534}{{\ttfamily arXiv:2012.08534
  [astro-ph.CO]}}.

\bibitem{Bernal:2016gxb}
J.~L. Bernal, L.~Verde, and A.~G. Riess, ``{The trouble with $H_0$},''
  \href{http://dx.doi.org/10.1088/1475-7516/2016/10/019}{{\em JCAP} {\bfseries
  1610} no.~10, (2016) 019},
\href{http://arxiv.org/abs/1607.05617}{{\ttfamily arXiv:1607.05617
  [astro-ph.CO]}}.
%%CITATION = ARXIV:1607.05617;%%.

\bibitem{Verde:2019ivm}
L.~Verde, T.~Treu, and A.~G. Riess,
  \href{http://dx.doi.org/10.1038/s41550-019-0902-0}{``{Tensions between the
  Early and the Late Universe},''} in {\em {Nature Astronomy 2019}}, vol.~3,
  p.~891.
\newblock 2019.
\newblock
\href{http://arxiv.org/abs/1907.10625}{{\ttfamily arXiv:1907.10625
  [astro-ph.CO]}}.
\newblock
%%CITATION = ARXIV:1907.10625;%%.

\bibitem{Knox:2019rjx}
L.~Knox and M.~Millea, ``{Hubble constant hunter’s guide},''
  \href{http://dx.doi.org/10.1103/PhysRevD.101.043533}{{\em Phys. Rev.}
  {\bfseries D101} no.~4, (2020) 043533},
\href{http://arxiv.org/abs/1908.03663}{{\ttfamily arXiv:1908.03663
  [astro-ph.CO]}}.
%%CITATION = ARXIV:1908.03663;%%.

\bibitem{Riess:2020sih}
A.~G. Riess, ``{The Expansion of the Universe is Faster than Expected},''
  \href{http://dx.doi.org/10.1038/s42254-019-0137-0}{{\em Nature Rev. Phys.}
  {\bfseries 2} no.~1, (2019) 10--12},
\href{http://arxiv.org/abs/2001.03624}{{\ttfamily arXiv:2001.03624
  [astro-ph.CO]}}.
%%CITATION = ARXIV:2001.03624;%%.

\bibitem{DiValentino:2020zio}
E.~Di~Valentino {\em et~al.}, ``{Snowmass2021 - Letter of interest cosmology
  intertwined II: The hubble constant tension},''
  \href{http://dx.doi.org/10.1016/j.astropartphys.2021.102605}{{\em Astropart.
  Phys.} {\bfseries 131} (2021) 102605},
  \href{http://arxiv.org/abs/2008.11284}{{\ttfamily arXiv:2008.11284
  [astro-ph.CO]}}.

\bibitem{DiValentino:2021izs}
E.~Di~Valentino, O.~Mena, S.~Pan, L.~Visinelli, W.~Yang, A.~Melchiorri, D.~F.
  Mota, A.~G. Riess, and J.~Silk, ``{In the realm of the Hubble
  tension\textemdash{}a review of solutions},''
  \href{http://dx.doi.org/10.1088/1361-6382/ac086d}{{\em Class. Quant. Grav.}
  {\bfseries 38} no.~15, (2021) 153001},
  \href{http://arxiv.org/abs/2103.01183}{{\ttfamily arXiv:2103.01183
  [astro-ph.CO]}}.

\bibitem{Perivolaropoulos:2021jda}
L.~Perivolaropoulos and F.~Skara, ``{Challenges for \ensuremath{\Lambda}CDM: An
  update},'' \href{http://dx.doi.org/10.1016/j.newar.2022.101659}{{\em New
  Astron. Rev.} {\bfseries 95} (2022) 101659},
  \href{http://arxiv.org/abs/2105.05208}{{\ttfamily arXiv:2105.05208
  [astro-ph.CO]}}.

\bibitem{Abdalla:2022yfr}
E.~Abdalla {\em et~al.}, ``{Cosmology intertwined: A review of the particle
  physics, astrophysics, and cosmology associated with the cosmological
  tensions and anomalies},''
  \href{http://dx.doi.org/10.1016/j.jheap.2022.04.002}{{\em JHEAp} {\bfseries
  34} (2022) 49--211}, \href{http://arxiv.org/abs/2203.06142}{{\ttfamily
  arXiv:2203.06142 [astro-ph.CO]}}.

\bibitem{Schoneberg:2021qvd}
N.~Sch\"oneberg, G.~F. Abell\'an, A.~P. S\'anchez, S.~J. Witte, c.~V. Poulin,
  and J.~Lesgourgues, ``{The $H_0$ Olympics: A fair ranking of proposed
  models},'' \href{http://arxiv.org/abs/2107.10291}{{\ttfamily arXiv:2107.10291
  [astro-ph.CO]}}.

\bibitem{Jedamzik:2020zmd}
K.~Jedamzik, L.~Pogosian, and G.-B. Zhao, ``{Why reducing the cosmic sound
  horizon alone can not fully resolve the Hubble tension},''
  \href{http://dx.doi.org/10.1038/s42005-021-00628-x}{{\em Commun. in Phys.}
  {\bfseries 4} (2021) 123}, \href{http://arxiv.org/abs/2010.04158}{{\ttfamily
  arXiv:2010.04158 [astro-ph.CO]}}.

\bibitem{Cai:2021weh}
R.-G. Cai, Z.-K. Guo, S.-J. Wang, W.-W. Yu, and Y.~Zhou, ``{No-go guide for the
  Hubble tension: Late-time solutions},''
  \href{http://dx.doi.org/10.1103/PhysRevD.105.L021301}{{\em Phys. Rev. D}
  {\bfseries 105} no.~2, (2022) L021301},
  \href{http://arxiv.org/abs/2107.13286}{{\ttfamily arXiv:2107.13286
  [astro-ph.CO]}}.

\bibitem{Cai:2022dkh}
R.-G. Cai, Z.-K. Guo, S.-J. Wang, W.-W. Yu, and Y.~Zhou, ``{No-go guide for
  late-time solutions to the Hubble tension: Matter perturbations},''
  \href{http://dx.doi.org/10.1103/PhysRevD.106.063519}{{\em Phys. Rev. D}
  {\bfseries 106} no.~6, (2022) 063519},
  \href{http://arxiv.org/abs/2202.12214}{{\ttfamily arXiv:2202.12214
  [astro-ph.CO]}}.

\bibitem{Riess:2021jrx}
A.~G. Riess {\em et~al.}, ``{A Comprehensive Measurement of the Local Value of
  the Hubble Constant with 1 km/s/Mpc Uncertainty from the Hubble Space
  Telescope and the SH0ES Team},'' {\em submitted to Astrophys. J.} (12, 2021)
  , \href{http://arxiv.org/abs/2112.04510}{{\ttfamily arXiv:2112.04510
  [astro-ph.CO]}}.

\bibitem{DES:2022tgg}
{\bfseries DES} Collaboration, C.~Meldorf {\em et~al.}, ``{The Dark Energy
  Survey Supernova Program results: Type Ia Supernova brightness correlates
  with host galaxy dust},'' \href{http://arxiv.org/abs/2206.06928}{{\ttfamily
  arXiv:2206.06928 [astro-ph.CO]}}.

\bibitem{Wojtak:2022bct}
R.~Wojtak and J.~Hjorth, ``{Intrinsic tension in the supernova sector of the
  local Hubble constant measurement and its implications},''
  \href{http://arxiv.org/abs/2206.08160}{{\ttfamily arXiv:2206.08160
  [astro-ph.CO]}}.

\bibitem{Rose:2022zmu}
B.~M. Rose, B.~Popovic, D.~Scolnic, and D.~Brout, ``{Constraining R$_V$
  Variation Using Highly Reddened Type Ia Supernovae from the Pantheon+
  Sample},'' \href{http://arxiv.org/abs/2206.09950}{{\ttfamily arXiv:2206.09950
  [astro-ph.CO]}}.

\bibitem{Dixon:2022ryo}
M.~Dixon {\em et~al.}, ``{Using Host Galaxy Spectroscopy to Explore Systematics
  in the Standardisation of Type Ia Supernovae},''
  \href{http://arxiv.org/abs/2206.12085}{{\ttfamily arXiv:2206.12085
  [astro-ph.CO]}}.

\bibitem{DES:2022qsy}
{\bfseries DES} Collaboration, P.~Wiseman {\em et~al.}, ``{A galaxy-driven
  model of type Ia supernova luminosity variations},''
  \href{http://arxiv.org/abs/2207.05583}{{\ttfamily arXiv:2207.05583
  [astro-ph.GA]}}.

\bibitem{DES:2022zpw}
{\bfseries DES} Collaboration, L.~Kelsey {\em et~al.}, ``{Concerning Colour:
  The Effect of Environment on Type Ia Supernova Colour in the Dark Energy
  Survey},'' \href{http://arxiv.org/abs/2208.01357}{{\ttfamily arXiv:2208.01357
  [astro-ph.CO]}}.

\bibitem{Jones:2022tsf}
D.~O. Jones, W.~D. Kenworthy, M.~Dai, R.~J. Foley, R.~Kessler, J.~D.~R. Pierel,
  and M.~R. Siebert, ``{A Spectroscopic Model of the Type Ia Supernova $-$ Host
  Galaxy Mass Correlation from SALT3},''
  \href{http://arxiv.org/abs/2209.05584}{{\ttfamily arXiv:2209.05584
  [astro-ph.CO]}}.

\bibitem{Kelly:2009iy}
P.~L. Kelly, M.~Hicken, D.~L. Burke, K.~S. Mandel, and R.~P. Kirshner,
  ``{Hubble Residuals of Nearby Type Ia Supernovae Are Correlated with Host
  Galaxy Masses},'' \href{http://dx.doi.org/10.1088/0004-637X/715/2/743}{{\em
  Astrophys. J.} {\bfseries 715} (2010) 743--756},
  \href{http://arxiv.org/abs/0912.0929}{{\ttfamily arXiv:0912.0929
  [astro-ph.CO]}}.

\bibitem{SNLS:2010kps}
{\bfseries SNLS} Collaboration, M.~Sullivan {\em et~al.}, ``{The Dependence of
  Type Ia Supernova Luminosities on their Host Galaxies},''
  \href{http://dx.doi.org/10.1111/j.1365-2966.2010.16731.x}{{\em Mon. Not. Roy.
  Astron. Soc.} {\bfseries 406} (2010) 782--802},
  \href{http://arxiv.org/abs/1003.5119}{{\ttfamily arXiv:1003.5119
  [astro-ph.CO]}}.

\bibitem{SDSS:2010swx}
{\bfseries SDSS} Collaboration, H.~Lampeitl {\em et~al.}, ``{The Effect of Host
  Galaxies on Type Ia Supernovae in the SDSS-II Supernova Survey},''
  \href{http://dx.doi.org/10.1088/0004-637X/722/1/566}{{\em Astrophys. J.}
  {\bfseries 722} (2010) 566--576},
  \href{http://arxiv.org/abs/1005.4687}{{\ttfamily arXiv:1005.4687
  [astro-ph.CO]}}.

\bibitem{Gupta:2011pa}
R.~R. Gupta {\em et~al.}, ``{Improved Constraints on Type Ia Supernova Host
  Galaxy Properties using Multi-Wavelength Photometry and their Correlations
  with Supernova Properties},''
  \href{http://dx.doi.org/10.1088/0004-637X/740/2/92}{{\em Astrophys. J.}
  {\bfseries 740} (2011) 92}, \href{http://arxiv.org/abs/1107.6003}{{\ttfamily
  arXiv:1107.6003 [astro-ph.CO]}}. [Erratum: Astrophys.J. 741, 127 (2011)].

\bibitem{Johansson:2012si}
J.~Johansson, D.~Thomas, J.~Pforr, C.~Maraston, R.~C. Nichol, M.~Smith,
  H.~Lampeitl, A.~Beifiori, R.~R. Gupta, and D.~P. Schneider, ``{SNe Ia host
  galaxy properties from Sloan Digital Sky Survey-II spectroscopy},''
  \href{http://dx.doi.org/10.1093/mnras/stt1408}{{\em Mon. Not. Roy. Astron.
  Soc.} {\bfseries 435} (2013) 1680},
  \href{http://arxiv.org/abs/1211.1386}{{\ttfamily arXiv:1211.1386
  [astro-ph.CO]}}.

\bibitem{Childress:2013xna}
M.~J. Childress {\em et~al.}, ``{Host Galaxy Properties and Hubble Residuals of
  Type Ia Supernovae from the Nearby Supernova Factory},''
  \href{http://dx.doi.org/10.1088/0004-637X/770/2/108}{{\em Astrophys. J.}
  {\bfseries 770} (2013) 108}, \href{http://arxiv.org/abs/1304.4720}{{\ttfamily
  arXiv:1304.4720 [astro-ph.CO]}}.

\bibitem{NearbySupernovafactory:2013qtg}
{\bfseries Nearby Supernova factory} Collaboration, M.~Rigault {\em et~al.},
  ``{Evidence of Environmental Dependencies of Type Ia Supernovae from the
  Nearby Supernova Factory indicated by Local H$\alpha$},''
  \href{http://dx.doi.org/10.1051/0004-6361/201322104}{{\em Astron. Astrophys.}
  {\bfseries 560} (2013) A66}, \href{http://arxiv.org/abs/1309.1182}{{\ttfamily
  arXiv:1309.1182 [astro-ph.CO]}}.

\bibitem{Rigault:2014kaa}
M.~Rigault {\em et~al.}, ``{Confirmation of a Star Formation Bias in Type Ia
  Supernova Distances and its Effect on Measurement of the Hubble Constant},''
  \href{http://dx.doi.org/10.1088/0004-637X/802/1/20}{{\em Astrophys. J.}
  {\bfseries 802} no.~1, (2015) 20},
  \href{http://arxiv.org/abs/1412.6501}{{\ttfamily arXiv:1412.6501
  [astro-ph.CO]}}.

\bibitem{Jones:2015uaa}
D.~O. Jones, A.~G. Riess, and D.~M. Scolnic, ``{Reconsidering the Effects of
  Local Star Formation On Type Ia Supernova Cosmology},''
  \href{http://dx.doi.org/10.1088/0004-637X/812/1/31}{{\em Astrophys. J.}
  {\bfseries 812} no.~1, (2015) 31},
  \href{http://arxiv.org/abs/1506.02637}{{\ttfamily arXiv:1506.02637
  [astro-ph.CO]}}.

\bibitem{Uddin:2017rmc}
S.~A. Uddin, J.~Mould, C.~Lidman, V.~Ruhlmann-Kleider, and B.~R. Zhang, ``{The
  influence of Host Galaxies in Type Ia Supernova Cosmology},''
  \href{http://dx.doi.org/10.3847/1538-4357/aa8df7}{{\em Astrophys. J.}
  {\bfseries 848} no.~1, (2017) 56},
  \href{http://arxiv.org/abs/1709.05830}{{\ttfamily arXiv:1709.05830
  [astro-ph.CO]}}.

\bibitem{Roman2018}
M.~{Roman}, D.~{Hardin}, M.~{Betoule}, P.~{Astier}, C.~{Balland}, R.~S.
  {Ellis}, S.~{Fabbro}, J.~{Guy}, I.~{Hook}, D.~A. {Howell}, C.~{Lidman},
  A.~{Mitra}, A.~{M{\"o}ller}, A.~M. {Mour{\~a}o}, J.~{Neveu},
  N.~{Palanque-Delabrouille}, C.~J. {Pritchet}, N.~{Regnault},
  V.~{Ruhlmann-Kleider}, C.~{Saunders}, and M.~{Sullivan}, ``{Dependence of
  Type Ia supernova luminosities on their local environment},''
  \href{http://dx.doi.org/10.1051/0004-6361/201731425}{{\em A\& A} {\bfseries
  615} (July, 2018) A68}, \href{http://arxiv.org/abs/1706.07697}{{\ttfamily
  arXiv:1706.07697 [astro-ph.GA]}}.

\bibitem{Jones:2018vbn}
D.~O. Jones {\em et~al.}, ``{Should Type Ia Supernova Distances be Corrected
  for their Local Environments?},''
  \href{http://dx.doi.org/10.3847/1538-4357/aae2b9}{{\em Astrophys. J.}
  {\bfseries 867} no.~2, (2018) 108},
  \href{http://arxiv.org/abs/1805.05911}{{\ttfamily arXiv:1805.05911
  [astro-ph.CO]}}.

\bibitem{Rose:2019ncv}
B.~M. Rose, P.~M. Garnavich, and M.~A. Berg, ``{Think Global, Act Local: The
  Influence of Environment Age and Host Mass on Type Ia Supernova Light
  Curves},'' \href{http://dx.doi.org/10.3847/1538-4357/ab0704}{{\em Astrophys.
  J.} {\bfseries 874} no.~1, (2019) 32},
  \href{http://arxiv.org/abs/1902.01433}{{\ttfamily arXiv:1902.01433
  [astro-ph.CO]}}.

\bibitem{Brout:2020msh}
D.~Brout and D.~Scolnic, ``{It\textquoteright{}s Dust: Solving the Mysteries of
  the Intrinsic Scatter and Host-galaxy Dependence of Standardized Type Ia
  Supernova Brightnesses},''
  \href{http://dx.doi.org/10.3847/1538-4357/abd69b}{{\em Astrophys. J.}
  {\bfseries 909} no.~1, (2021) 26},
  \href{http://arxiv.org/abs/2004.10206}{{\ttfamily arXiv:2004.10206
  [astro-ph.CO]}}.

\bibitem{Popovic:2021cwq}
B.~Popovic, D.~Brout, R.~Kessler, D.~Scolnic, and L.~Lu, ``{Improved Treatment
  of Host-Galaxy Correlations in Cosmological Analyses With Type Ia
  Supernovae},'' \href{http://dx.doi.org/10.3847/1538-4357/abf14f}{{\em
  Astrophys. J.} {\bfseries 913} no.~1, (2021) 49},
  \href{http://arxiv.org/abs/2102.01776}{{\ttfamily arXiv:2102.01776
  [astro-ph.CO]}}.

\bibitem{1992AJ....103.1427T}
E.~L. {Turner}, R.~{Cen}, and J.~P. {Ostriker}, ``{The Relationship of Local
  measures of Hubble's Constant to its Global Value},''
  \href{http://dx.doi.org/10.1086/116156}{{\em Astronomical Journal} {\bfseries
  103} (May, 1992) 1427}.

\bibitem{Shi:1995nq}
X.~Shi, L.~M. Widrow, and L.~J. Dursi, ``{Measuring hubble's constant in our
  inhomogeneous universe},''
  \href{http://dx.doi.org/10.1093/mnras/281.2.565}{{\em Mon. Not. Roy. Astron.
  Soc.} {\bfseries 281} (1996) 565},
  \href{http://arxiv.org/abs/astro-ph/9506120}{{\ttfamily
  arXiv:astro-ph/9506120}}.

\bibitem{Shi:1997aa}
X.-D. Shi and M.~S. Turner, ``{Expectations for the difference between local
  and global measurements of the Hubble constant},''
  \href{http://dx.doi.org/10.1086/305169}{{\em Astrophys. J.} {\bfseries 493}
  (1998) 519}, \href{http://arxiv.org/abs/astro-ph/9707101}{{\ttfamily
  arXiv:astro-ph/9707101}}.

\bibitem{Wang:1997tp}
Y.~Wang, D.~N. Spergel, and E.~L. Turner, ``{Implications of cosmic microwave
  background anisotropies for large scale variations in Hubble's constant},''
  \href{http://dx.doi.org/10.1086/305539}{{\em Astrophys. J.} {\bfseries 498}
  (1998) 1}, \href{http://arxiv.org/abs/astro-ph/9708014}{{\ttfamily
  arXiv:astro-ph/9708014}}.

\bibitem{Sinclair:2010sb}
B.~Sinclair, T.~M. Davis, and T.~Haugbolle, ``{Residual Hubble-bubble effects
  on supernova cosmology},''
  \href{http://dx.doi.org/10.1088/0004-637X/718/2/1445}{{\em Astrophys. J.}
  {\bfseries 718} (2010) 1445--1455},
  \href{http://arxiv.org/abs/1006.0911}{{\ttfamily arXiv:1006.0911
  [astro-ph.CO]}}.

\bibitem{Marra:2013rba}
V.~Marra, L.~Amendola, I.~Sawicki, and W.~Valkenburg, ``{Cosmic variance and
  the measurement of the local Hubble parameter},''
  \href{http://dx.doi.org/10.1103/PhysRevLett.110.241305}{{\em Phys. Rev.
  Lett.} {\bfseries 110} no.~24, (2013) 241305},
  \href{http://arxiv.org/abs/1303.3121}{{\ttfamily arXiv:1303.3121
  [astro-ph.CO]}}.

\bibitem{Ben-Dayan:2014swa}
I.~Ben-Dayan, R.~Durrer, G.~Marozzi, and D.~J. Schwarz, ``{The value of $H_0$
  in the inhomogeneous Universe},''
  \href{http://dx.doi.org/10.1103/PhysRevLett.112.221301}{{\em Phys. Rev.
  Lett.} {\bfseries 112} (2014) 221301},
  \href{http://arxiv.org/abs/1401.7973}{{\ttfamily arXiv:1401.7973
  [astro-ph.CO]}}.

\bibitem{Camarena:2018nbr}
D.~Camarena and V.~Marra, ``{Impact of the cosmic variance on $H_0$ on
  cosmological analyses},''
  \href{http://dx.doi.org/10.1103/PhysRevD.98.023537}{{\em Phys. Rev. D}
  {\bfseries 98} no.~2, (2018) 023537},
  \href{http://arxiv.org/abs/1805.09900}{{\ttfamily arXiv:1805.09900
  [astro-ph.CO]}}.

\bibitem{Wojtak:2013gda}
R.~Wojtak, A.~Knebe, W.~A. Watson, I.~T. Iliev, S.~Heß, D.~Rapetti, G.~Yepes,
  and S.~Gottlöber, ``{Cosmic variance of the local Hubble flow in large-scale
  cosmological simulations},''
  \href{http://dx.doi.org/10.1093/mnras/stt2321}{{\em Mon. Not. Roy. Astron.
  Soc.} {\bfseries 438} no.~2, (2014) 1805--1812},
\href{http://arxiv.org/abs/1312.0276}{{\ttfamily arXiv:1312.0276
  [astro-ph.CO]}}.
%%CITATION = ARXIV:1312.0276;%%.

\bibitem{Odderskov:2014hqa}
I.~Odderskov, S.~Hannestad, and T.~Haugbølle, ``{On the local variation of the
  Hubble constant},''
  \href{http://dx.doi.org/10.1088/1475-7516/2014/10/028}{{\em JCAP} {\bfseries
  1410} (2014) 028},
\href{http://arxiv.org/abs/1407.7364}{{\ttfamily arXiv:1407.7364
  [astro-ph.CO]}}.
%%CITATION = ARXIV:1407.7364;%%.

\bibitem{Wu:2017fpr}
H.-Y. Wu and D.~Huterer, ``{Sample variance in the local measurements of the
  Hubble constant},'' \href{http://dx.doi.org/10.1093/mnras/stx1967}{{\em Mon.
  Not. Roy. Astron. Soc.} {\bfseries 471} no.~4, (2017) 4946--4955},
\href{http://arxiv.org/abs/1706.09723}{{\ttfamily arXiv:1706.09723
  [astro-ph.CO]}}.
%%CITATION = ARXIV:1706.09723;%%.

\bibitem{Kenworthy:2019qwq}
W.~D. Kenworthy, D.~Scolnic, and A.~Riess, ``{The Local Perspective on the
  Hubble Tension: Local Structure Does Not Impact Measurement of the Hubble
  Constant},'' \href{http://dx.doi.org/10.3847/1538-4357/ab0ebf}{{\em
  Astrophys. J.} {\bfseries 875} no.~2, (2019) 145},
  \href{http://arxiv.org/abs/1901.08681}{{\ttfamily arXiv:1901.08681
  [astro-ph.CO]}}.

\bibitem{Lukovic:2019ryg}
V.~V. Luković, B.~S. Haridasu, and N.~Vittorio, ``{Exploring the evidence for
  a large local void with supernovae Ia data},''
  \href{http://dx.doi.org/10.1093/mnras/stz3070}{{\em Mon. Not. Roy. Astron.
  Soc.} {\bfseries 491} no.~2, (2020) 2075--2087},
\href{http://arxiv.org/abs/1907.11219}{{\ttfamily arXiv:1907.11219
  [astro-ph.CO]}}.
%%CITATION = ARXIV:1907.11219;%%.

\bibitem{Cai:2020tpy}
R.-G. Cai, J.-F. Ding, Z.-K. Guo, S.-J. Wang, and W.-W. Yu, ``{Do the
  observational data favor a local void?},''
  \href{http://dx.doi.org/10.1103/PhysRevD.103.123539}{{\em Phys. Rev. D}
  {\bfseries 103} no.~12, (2021) 123539},
  \href{http://arxiv.org/abs/2012.08292}{{\ttfamily arXiv:2012.08292
  [astro-ph.CO]}}.

\bibitem{Castello:2021uad}
S.~Castello, M.~H\"og\r{a}s, and E.~M\"ortsell, ``{A cosmological underdensity
  does not solve the Hubble tension},''
  \href{http://dx.doi.org/10.1088/1475-7516/2022/07/003}{{\em JCAP} {\bfseries
  07} no.~07, (2022) 003}, \href{http://arxiv.org/abs/2110.04226}{{\ttfamily
  arXiv:2110.04226 [astro-ph.CO]}}.

\bibitem{Camarena:2022iae}
D.~Camarena, V.~Marra, Z.~Sakr, and C.~Clarkson, ``{A void in the Hubble
  tension? The end of the line for the Hubble bubble},''
  \href{http://dx.doi.org/10.1088/1361-6382/ac8635}{{\em Class. Quant. Grav.}
  {\bfseries 39} no.~18, (2022) 184001},
  \href{http://arxiv.org/abs/2205.05422}{{\ttfamily arXiv:2205.05422
  [astro-ph.CO]}}.

\bibitem{Sheth:2000ii}
R.~K. Sheth and A.~Diaferio, ``{Peculiar velocities of galaxies and
  clusters},'' \href{http://dx.doi.org/10.1046/j.1365-8711.2001.04202.x}{{\em
  Mon. Not. Roy. Astron. Soc.} {\bfseries 322} (2001) 901},
  \href{http://arxiv.org/abs/astro-ph/0009166}{{\ttfamily
  arXiv:astro-ph/0009166}}.

\bibitem{Scolnic:2017caz}
D.~M. Scolnic {\em et~al.}, ``{The Complete Light-curve Sample of
  Spectroscopically Confirmed SNe Ia from Pan-STARRS1 and Cosmological
  Constraints from the Combined Pantheon Sample},''
  \href{http://dx.doi.org/10.3847/1538-4357/aab9bb}{{\em Astrophys. J.}
  {\bfseries 859} no.~2, (2018) 101},
  \href{http://arxiv.org/abs/1710.00845}{{\ttfamily arXiv:1710.00845
  [astro-ph.CO]}}.

\bibitem{Jones:2017udy}
D.~O. Jones {\em et~al.}, ``{Measuring Dark Energy Properties with
  Photometrically Classified Pan-STARRS Supernovae. II. Cosmological
  Parameters},'' \href{http://dx.doi.org/10.3847/1538-4357/aab6b1}{{\em
  Astrophys. J.} {\bfseries 857} no.~1, (2018) 51},
\href{http://arxiv.org/abs/1710.00846}{{\ttfamily arXiv:1710.00846
  [astro-ph.CO]}}.
%%CITATION = ARXIV:1710.00846;%%.

\bibitem{Scolnic:2021amr}
D.~Scolnic {\em et~al.}, ``{The Pantheon+ Analysis: The Full Dataset and
  Light-Curve Release},'' \href{http://arxiv.org/abs/2112.03863}{{\ttfamily
  arXiv:2112.03863 [astro-ph.CO]}}.

\bibitem{Brout:2022vxf}
D.~Brout {\em et~al.}, ``{The Pantheon+ Analysis: Cosmological Constraints},''
  \href{http://arxiv.org/abs/2202.04077}{{\ttfamily arXiv:2202.04077
  [astro-ph.CO]}}.

\bibitem{Peterson:2021hel}
E.~R. Peterson {\em et~al.}, ``{The Pantheon+ Analysis: Evaluating Peculiar
  Velocity Corrections in Cosmological Analyses with Nearby Type Ia
  Supernovae},'' \href{http://arxiv.org/abs/2110.03487}{{\ttfamily
  arXiv:2110.03487 [astro-ph.CO]}}.

\bibitem{Foreman-Mackey:2012any}
D.~Foreman-Mackey, D.~W. Hogg, D.~Lang, and J.~Goodman, ``{emcee: The MCMC
  Hammer},'' \href{http://dx.doi.org/10.1086/670067}{{\em Publ. Astron. Soc.
  Pac.} {\bfseries 125} (2013) 306--312},
  \href{http://arxiv.org/abs/1202.3665}{{\ttfamily arXiv:1202.3665
  [astro-ph.IM]}}.

\bibitem{Note1}
we have omitted 46 SNe Ia with their hosts too faint for survey depth so that
  their host masses are simply assigned in the lowest mass bin.

\bibitem{Scrimgeour:2012wt}
M.~Scrimgeour {\em et~al.}, ``{The WiggleZ Dark Energy Survey: the transition
  to large-scale cosmic homogeneity},''
  \href{http://dx.doi.org/10.1111/j.1365-2966.2012.21402.x}{{\em Mon. Not. Roy.
  Astron. Soc.} {\bfseries 425} (2012) 116--134},
  \href{http://arxiv.org/abs/1205.6812}{{\ttfamily arXiv:1205.6812
  [astro-ph.CO]}}.

\bibitem{footnote}
See the supplemental material for a theoretical estimation from the
  $\Lambda$CDM model on the variations in the measured local Hubble constants
  from an arbitrary discrete sample of distant SNe Ia with the same ambient
  density contrast estimated at a given scale as well as the statistical
  analysis for estimating the deviation significance of the observational
  non-local slope with respect to the $\Lambda$CDM expectation.

\bibitem{Lavaux:2019fjr}
G.~Lavaux, J.~Jasche, and F.~Leclercq, ``{Systematic-free inference of the
  cosmic matter density field from SDSS3-BOSS data},''
\href{http://arxiv.org/abs/1909.06396}{{\ttfamily arXiv:1909.06396
  [astro-ph.CO]}}.
%%CITATION = ARXIV:1909.06396;%%.

\bibitem{SDSS-III:2015hof}
{\bfseries SDSS-III} Collaboration, S.~Alam {\em et~al.}, ``{The Eleventh and
  Twelfth Data Releases of the Sloan Digital Sky Survey: Final Data from
  SDSS-III},'' \href{http://dx.doi.org/10.1088/0067-0049/219/1/12}{{\em
  Astrophys. J. Suppl.} {\bfseries 219} no.~1, (2015) 12},
  \href{http://arxiv.org/abs/1501.00963}{{\ttfamily arXiv:1501.00963
  [astro-ph.IM]}}.

\bibitem{BOSS:2016wmc}
{\bfseries BOSS} Collaboration, S.~Alam {\em et~al.}, ``{The clustering of
  galaxies in the completed SDSS-III Baryon Oscillation Spectroscopic Survey:
  cosmological analysis of the DR12 galaxy sample},''
  \href{http://dx.doi.org/10.1093/mnras/stx721}{{\em Mon. Not. Roy. Astron.
  Soc.} {\bfseries 470} no.~3, (2017) 2617--2652},
  \href{http://arxiv.org/abs/1607.03155}{{\ttfamily arXiv:1607.03155
  [astro-ph.CO]}}.

\bibitem{Cai:2021wgv}
R.-G. Cai, Z.-K. Guo, L.~Li, S.-J. Wang, and W.-W. Yu, ``{Chameleon dark energy
  can resolve the Hubble tension},''
  \href{http://dx.doi.org/10.1103/PhysRevD.103.L121302}{{\em Phys. Rev. D}
  {\bfseries 103} no.~12, (2021) L121302},
  \href{http://arxiv.org/abs/2102.02020}{{\ttfamily arXiv:2102.02020
  [astro-ph.CO]}}.

\bibitem{1980lssu.book.....P}
P.~J.~E. {Peebles}, {\em {The large-scale structure of the universe}}.
\newblock 1980.

\bibitem{1993ppc..book.....P}
P.~J.~E. {Peebles}, {\em {Principles of Physical Cosmology}}.
\newblock 1993.

\bibitem{Hogg:2004vw}
D.~W. Hogg, D.~J. Eisenstein, M.~R. Blanton, N.~A. Bahcall, J.~Brinkmann, J.~E.
  Gunn, and D.~P. Schneider, ``{Cosmic homogeneity demonstrated with luminous
  red galaxies},'' \href{http://dx.doi.org/10.1086/429084}{{\em Astrophys. J.}
  {\bfseries 624} (2005) 54--58},
  \href{http://arxiv.org/abs/astro-ph/0411197}{{\ttfamily
  arXiv:astro-ph/0411197}}.

\bibitem{Hartlap:2006kj}
J.~Hartlap, P.~Simon, and P.~Schneider, ``{Why your model parameter confidences
  might be too optimistic: Unbiased estimation of the inverse covariance
  matrix},'' \href{http://dx.doi.org/10.1051/0004-6361:20066170}{{\em Astron.
  Astrophys.} {\bfseries 464} (2007) 399},
  \href{http://arxiv.org/abs/astro-ph/0608064}{{\ttfamily
  arXiv:astro-ph/0608064}}.

\end{thebibliography}\endgroup

\end{document}